\documentclass[aps,pre,superscriptaddress,showpacs,floatfix,twocolumn]{revtex4-1}

\usepackage{graphicx}   
\usepackage{amssymb}
\begin{document}

\title{A dynamical network model for age-related health deficits and mortality} 
\author{Swadhin Taneja}	
\email{swadhin.taneja@dal.ca}	
\affiliation{Dept. of Physics and Atmospheric Science, Dalhousie University, Halifax, Nova Scotia, Canada B3H 4R2}
\affiliation{Department of Medicine, Dalhousie University, Halifax, Nova Scotia, Canada  B3H 2Y9}
\affiliation{Division of Geriatric Medicine, Dalhousie University , Halifax, Nova Scotia, Canada B3H 2E1}
\author{Arnold B. Mitnitski}	
\email{arnold.mitnitski@dal.ca}
\affiliation{Department of Medicine, Dalhousie University, Halifax, Nova Scotia, Canada  B3H 2Y9}
\author{Kenneth Rockwood}	
\email{kenneth.rockwood@dal.ca}
\affiliation{Department of Medicine, Dalhousie University, Halifax, Nova Scotia, Canada  B3H 2Y9}
\affiliation{Division of Geriatric Medicine, Dalhousie University , Halifax, Nova Scotia, Canada B3H 2E1}
\author{Andrew D. Rutenberg}	
\email{andrew.rutenberg@dal.ca}
\affiliation{Dept. of Physics and Atmospheric Science, Dalhousie University, Halifax, Nova Scotia, Canada B3H 4R2}

\date{\today}
\begin{abstract}  
How long people live depends on their health, and how it changes with age. Individual health can be tracked by the accumulation of age-related health deficits. The fraction of age-related deficits is a simple quantitative measure of human aging. This quantitative frailty index ($F$) is as good as chronological age in predicting  mortality.  In this paper, we use a dynamical network model of deficits to explore the effects of interactions between deficits, deficit damage and repair processes, and the connection between the $F$ and mortality.  With our model, we qualitatively reproduce Gompertz's law of increasing human mortality with age, the broadening of the $F$ distribution with age,  the characteristic non-linear increase of the $F$ with age, and the increased mortality of high-frailty individuals.  No explicit time-dependence in damage or repair rates is needed in our model. Instead, implicit time-dependence arises through deficit interactions -- so that the average deficit damage rates increases, and deficit repair rates decreases, with age .  We use a simple mortality criterion, where mortality occurs when the most connected node is damaged. 
\end{abstract}
\maketitle

\section{Introduction} 

Average human mortality rates increase exponentially with age, following Gompertz's law \cite{gompertz}. This exponential growth holds well for ages between 40 and 100 years, and possibly older \cite{beyondgompertz}. Individual mortality is preceded by the dynamical process of {\it aging}, which can be viewed as the accumulation of organismal damage with time \cite{accumdamage}. Because individuals can accumulate health problems at different times, the health status of populations of a given age is heterogeneous. Individual ``frailty'' could account for at least some of the differences in mortality rates for individuals of the same age \cite{Vaupel1979}, where frailty results from the individual accumulation of damage during aging.

One quantitative measure of frailty is the time-dependent ``frailty index"  (FI) \cite{Mitnitski2001, Rockwood2002, Mitnitski2013, Mitnitski2015, Kulminski2007, Yashin2008}, quantitatively denoted $F$. Clinically, the frailty index is assessed by scoring a broad portfolio of possible age-related deficits as either $0$ (healthy) or $1$ (damaged). Then $F$ is the proportion that are damaged, and ranges from $0$ for perfectly healthy individuals to a theoretical maximum of $1$ \cite{Rockwood2002}. Remarkably, in elderly populations the frailty index is as predictive as chronological age for various health-outcomes -- including mortality \cite{Mitnitski2005, Kulminski2007b, Yashin2008}. Many different portfolios of age-related deficits, of different sizes, have similar behavior \cite{health, Gu2009}.  For example, an FI can be created from biomarkers and laboratory tests \cite{biochem}. An FI can also be calculated for mice, and shows similar behavior as for humans \cite{mice}. Nevertheless, our understanding of frailty accumulation in individuals remains largely empirical \cite{Mitnitski2013}.

The distribution of the frailty index broadens significantly with age \cite{Rockwood2004, Gu2009}.  This indicates that the stochasticity of frailty dynamics is significant \cite{Vaupel1979, Yashin2012}. Indeed, longitudinal studies show both increases and decreases of $F$ with increasing age of individuals  \cite{Mitnitski2006, Gill2006, Mitnitski2012}, and there are also direct indications that at least some individual deficits are reversible \cite{Mitnitski2015, Mitnitski2006, reversible}.  Two additional observations focus our approach. First, the time-dependent average $F(t)$ shows a significant upward curvature vs. age \cite{Mitnitski2001, Mitnitski2002, Mitnitski2005, Gu2009, Mitnitski2013, Theou2014, health}. While linear accumulation of deficits with age indicates independent damage processes for each deficit, and negative curvature can indicate saturation, upward curvature indicates an increasing rate of deficit accumulation with age \cite{Kulminski2007c}.   Second, average mortality rates increase with increasing $F$ \cite{Mitnitski2006}.  This increase underlies the additional clinical predictive value of the FI, as compared to age alone. 

Early quantitative models of human mortality, such as that of Strehler and Mildvan \cite{Strehler1960}, focused on obtaining the time-dependent mortality directly. More recently, to capture the upward curvature of $F(t)$, researchers have introduced explicit time-dependent damage or repair rates in the dynamics of individual health deficits \cite{Mitnitski2013, Mitnitski2015}. To relate deficits with mortality, researchers have also included explicitly time-dependent dynamics for individual deficits \cite{Arbeev2011, Yashin2012}.

Conceptually, individuals are comprised of interconnected, or networked, processes and components. Networks underly both healthy function \cite{west} and  human disease \cite{networkdisease}. It has been proposed that a network approach would be appropriate in the study of human aging at various scales \cite{Simko2009}. Indeed, earlier work proposed biochemical networks at the cellular level \cite{Kowald1996}, and   demonstrated networks of correlations between age-related deficits \cite{Rockwood2002b}.  Recently, Vural et al have presented a network model of animal mortality \cite{Vural2014}. 

Often complex networks are assumed to be scale-free, where the number of connections (edges) $k$ from each node has a power law distribution $k^{-\alpha}$ \cite{Barabasi}. Such scale-free networks are heterogeneous -- with a few large hubs that have a large numbers of connections, and many more nodes that are not well connected. Conversely, nodes in random networks are homogeneous --- all have similar connectivity. There are indications that networks of human physiology \cite{west}, disease \cite{Menche2015}, longevity-related proteins \cite{Wolfson2009}, and  age-related deficits \cite{Rockwood2002b} are heterogeneous, and may be scale-free \cite{west, Wolfson2009}. 
 
In order to better understand how deficit dynamics can lead to both the observed behavior of $F(t)$ and the observed mortality, we have developed a stochastic  model of interacting deficits. We represent an individual as an undirected network of connected deficit nodes, each of which has two stable states corresponding to healthy or damaged. In our model, damage and repair rates have no explicit time-dependence, but do depend on the state of connected nodes. 

For scale-free and random network models, random removal of more than a threshold fraction of nodes leads to a  failure of network connectivity  \cite{Barabasi, Albert2000}. To determine animal mortality, the network model of Vural {\em et al.} used a frailty threshold. However an explicit threshold on $F$ cannot recover the continuously increasing mortality rates with increasing $F$ that are seen with clinical human data \cite{Mitnitski2006}. To attempt to recover this behavior, we take mortality to occur when one (or more) of the most highly connected nodes is damaged, while frailty is represented by the average damage state of highly connected nodes that are not mortality nodes.

With our model we recover much of the existing frailty phenomenology, along with a Gompertz-like increase of mortality with age. This addresses two basic questions. First, can the upward curvature of $F(t)$ be captured {\em implicitly} by interactions between deficits, or must it imply an {\em explicit} age-dependence in, e.g., the damage or repair rates of individual deficits? Second, can the increase of mortality with respect to $F$ be captured implicitly by interacting deficits, or must it imply an explicit dynamical dependence of mortality on $F$? In both cases, we show that the behavior can arise naturally and implicitly from interacting deficits alone.

\begin{figure}[t*]   
\includegraphics[trim = 45mm 10mm 20mm 40mm, clip, width=0.5\textwidth]{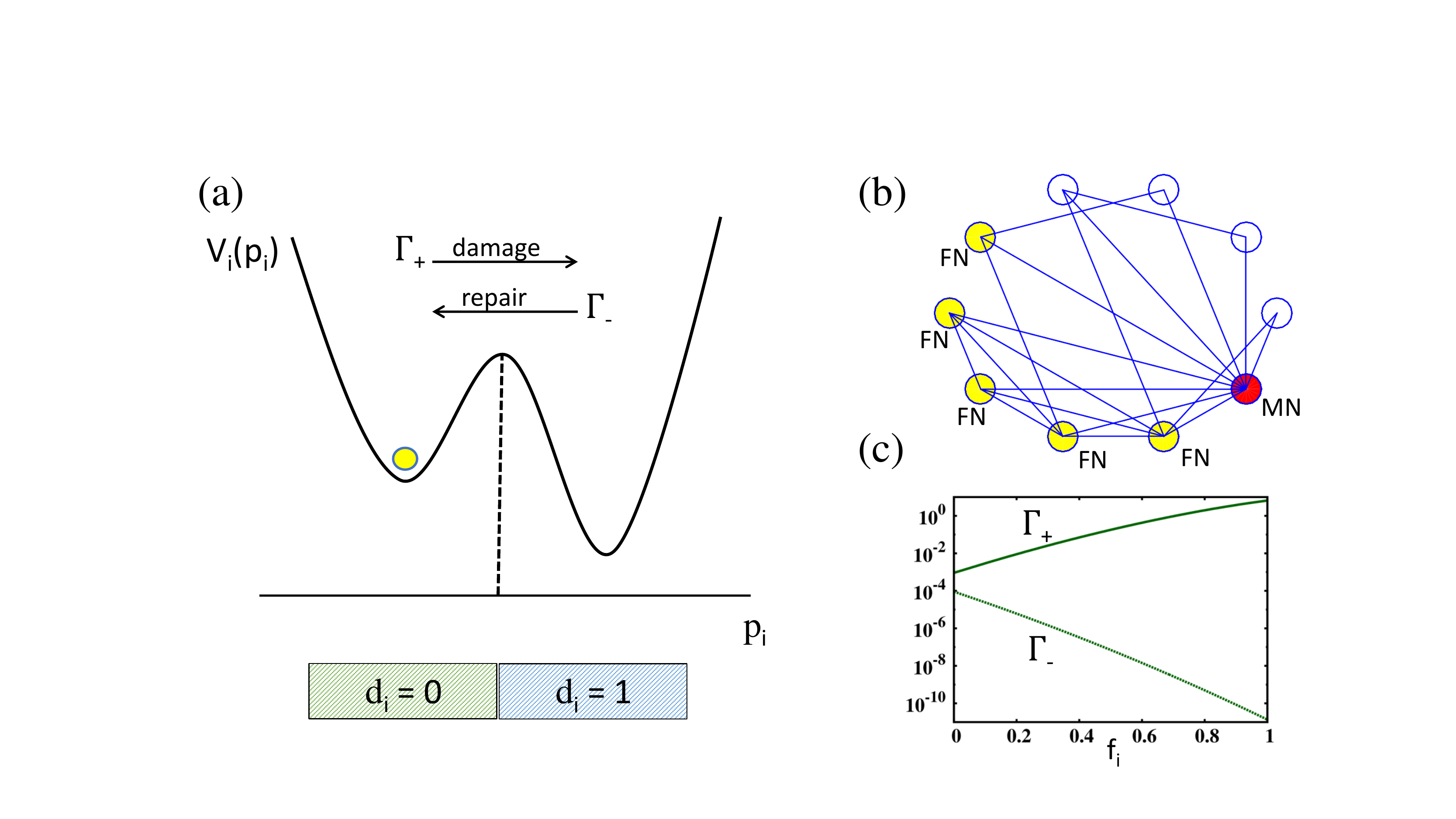}
\caption{(color online) (a) Transition rates $\Gamma_\pm$ between discrete undamaged ($d_i=0$) and damaged ($d_i=1$) states of a deficit node $i$ are motivated by stochastic dynamics of an asymmetric double well $\tilde{V}(p_i)$ with a continuous damage variable $p_i$, as illustrated.  (b) A small network of interactions (blue lines) between deficit nodes (circles) is illustrated. The most connected node  determines mortality (red circle, ``MN''), while the next most connected nodes determine frailty (yellow circles, ``FN''). (c) Damage ($\Gamma_+$) and repair ($\Gamma_-$) rates (per year), as indicated, for nodes with local frailty $f_i$.} 
\label{f:DoubleWell}
\end{figure}

\section{Model} 
For each individual, we consider  network nodes $i$ to represent continuous valued  physiological parameters $p_i(t)$, where increasing damage corresponds to increasing $p_i$, that each follow stochastic over-damped dynamics following an effective potential:
\begin{equation}  
	\frac{d  p_i}{dt}= - \frac{dV}{dp} (p_i)+\sigma f_i +\xi_i (t) 
\label{e:dynamics}
\end{equation}  
where $\xi_i (t)$ is Gaussian white noise, $\sigma$ is the interaction parameter, and $f_i$ is the average damage of all nodes connected with the $i$th node (i.e. a local frailty).  For simplicity, we take $V(p)$ to be an asymmetric double-well potential (see Fig.~\ref{f:DoubleWell})
\begin{equation}
	V(p)= V_0 \left[\frac{p^4}{4}-\frac{p^3}{3}(1+\theta) + \frac{p^2}{2}\theta \right]
\label{v}
\end{equation} 
with local minima at $p=0$ and $1$, and an intervening barrier at $\theta \approx 0.5$. The interaction term in Eqn.~\ref{e:dynamics} can also be represented by a tilted effective potential $\tilde{V}_i \equiv V(p_i)- \sigma p_i f_i$, which then shifts the local minima for each node and also shifts the intervening barrier $\theta_i$.

Under the effects of noisy dynamics, individual nodes will transition randomly between the two local minima corresponding to healthy states (with $p_i < \theta_i$) and damaged states (with $p_i > \theta_i$), which we discretely label with deficit indices $d_i=0$ and $d_i=1$, respectively.  The local frailty $f_i$ in Eqn.~\ref{e:dynamics} is then defined by $f_i \equiv \sum d_j/k_i$, where the sum is over the $k_i$ nodes connected with the $i$th node so that $f_i \in [0,1]$.  We can approximate the transition rates with Kramer's limiting rates \cite{Kramers1940},
\begin{equation}
\label{e:kramerrate}
	\Gamma_\pm(f_i)=\Gamma_0 ~exp \left(- \frac{{\Delta \tilde{V}} _\pm(f_i)}{D} \right),
\end{equation}
where ${\Delta \tilde{V}} _\pm$ are the barrier heights from the healthy or damaged sides, corresponding to damage $\Gamma_+$ and repair $\Gamma_-$ rates, respectively.  The interaction strength $\sigma$ is kept small enough so that local minima are well defined even when $f_i=1$.    Rates  depend on the local frailty $f_i$ through Eqn.~\ref{e:dynamics}. These transition rates, $\Gamma_\pm$ vs $f_i$, are illustrated in Fig.~\ref{f:DoubleWell}(c). The  rates show approximately exponential dependence on the local frailty.

While the asymmetric double well and stochastic transitions in Eqns.~\ref{e:dynamics} and \ref{v} are useful to understand our model, for purposes of computational efficiency we directly implement the Kramer's rates in Eqn.~\ref{e:kramerrate} for each node $i$ using exact random sampling of transition times between healthy and damaged states $d_i=0$ and $d_i=1$  \cite{Gillespie1977}.  After each transition of any node, local frailties $f_j$ and transition rates are updated for all connected nodes. 

Rather than determining the Kramer's rate parameters $\Gamma_0$ and $D$ directly from  Eqns.~\ref{e:dynamics} and \ref{v}, we tune them phenomenologically.  $\Gamma_0$ is adjusted so that our model results agree with the most likely death age in human mortality statistics, which is 87.5 years \cite{USAdata}. This amounts to a rescaling of time, or an overall rescaling of either side of Eqn.~\ref{e:dynamics}.  $V_0/D$ is adjusted to obtain a fixed asymmetry ratio of damage to repair rates, $R \equiv \Gamma_+(0)/\Gamma_-(0)$. This amounts to rescaling the noise amplitude $\xi$ and/or the potential $V_0$.

Each individual is represented by $N$ nodes, which are all initially undamaged at age $t=0$. The undirected network of interactions is generated for each individual  with a characteristic scale-free exponent $\alpha=3$ and average connectivity $\langle k \rangle$ \cite{Barabasi}.  For comparison (see appendix), random networks with $N$ nodes and average connectivity $\langle k \rangle$ were generated by connecting ${N \langle k \rangle}/{2}$ edges between randomly selected pairs of nodes. At least $10000$ individuals are sampled for each parameter set in the paper.  Unless otherwise noted, we use standard model parameters of $N=800$ nodes, average connectivity $\langle k \rangle = 10$, potential shape $\theta=0.48$, and scaled interaction strength $\tilde{\sigma} = \sigma/D = 25$, and asymmetry ratio $R = 10$.   The dependence of our results on these parameter choices are explored below. 

In most of this paper, and unless otherwise noted, we have defined mortality to occur when the most connected node is damaged.  This is an attractively simple mortality condition. We have also explored using more than one of the most connected nodes to define mortality (see Fig.~\ref{f:OR} below, or  Figs.~\ref{OR} and \ref{AND}). When there is degeneracy, we choose mortality nodes at random from among the most connected. We define the frailty index to be
\begin{equation}
	F_n(t) = \sum_j d_j(t)/n,
\end{equation}
where the sum is over the $n$ most highly connected {\em non-mortality} nodes.  Since each $d_j \in \{0,1\}$, we have $F \in [0,1]$. Unless otherwise noted, we take $n=30$.   $F_n$ is a diagnostic measure, and is not directly involved in either damage or repair of individual nodes or in mortality.  This allows us to critically examine the relationship between $F(t)$ and mortality without explicit overlap between the two. Our focus in this paper is on older individuals, above the age of 50.

\begin{figure}[t*]   
\centering
\begin{minipage}{.45\textwidth}
\includegraphics[trim = 15mm 10mm 5mm 10mm, clip, width=\textwidth]{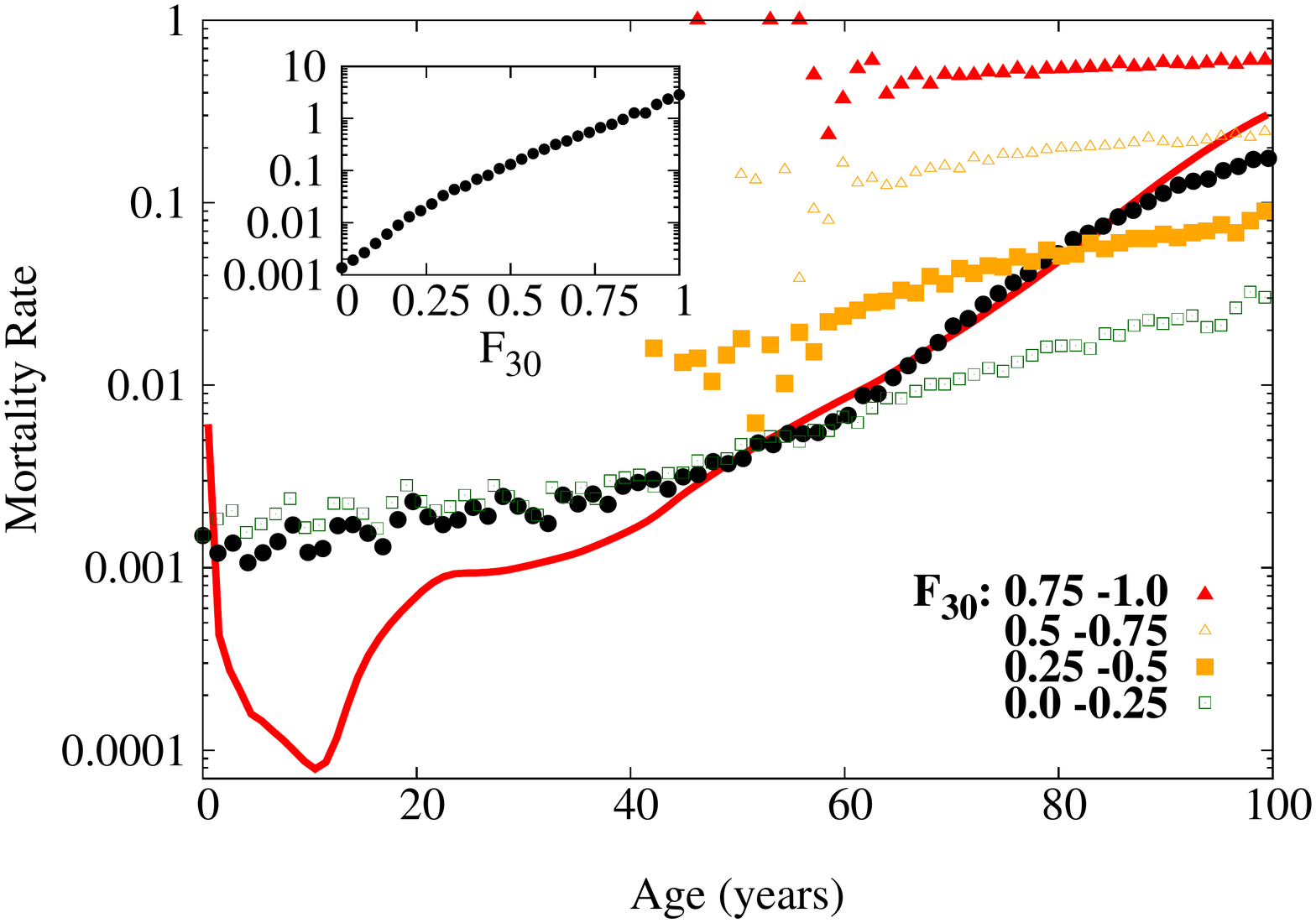}
\caption{(color online) The mortality rate, or chance of death per year, vs age from 2010 U.S. population life tables \cite{USAdata} is shown as a solid red line. Our model mortality rates are shown with filled black circles, using default parameter values.  We have qualitative agreement for ages above 50. We also show, as indicated by the legend, mortality rates for four quintiles of frailty index ($F_{30}$) vs age. In the inset we show model mortality rates vs $F_{30}$.}
\label{f:Gompertzdata}

\end{minipage} \quad
\begin{minipage}{.45\textwidth} 
\includegraphics[trim = 15mm 5mm 5mm 10mm, clip, width=\textwidth]{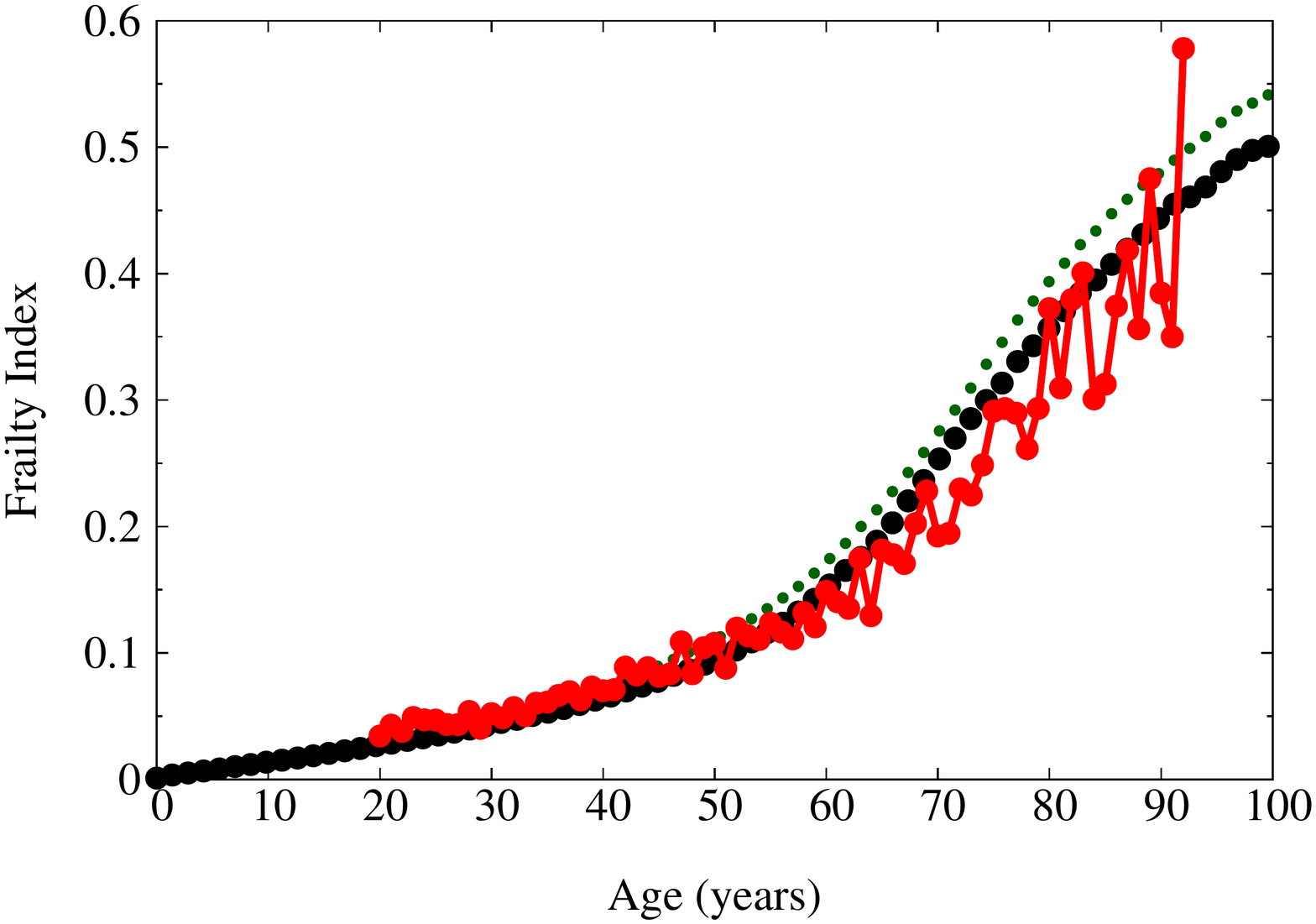}
\caption{(color online) The frailty index $F$ vs age determined from Canadian population data \cite{Mitnitski2013} is shown with red circles and a red line. Our model results $F_n$ are also shown, with $n=30$ and $n=799$ by black and small green circles, respectively.}
\label{f:FIdata}
\end{minipage}
\end{figure}

\section{Results} 
The solid red line in Fig.~\ref{f:Gompertzdata} shows United States mortality rate statistics (chance of death per year) vs. age \cite{USAdata}. The approximately exponential growth of mortality after age 50 is Gompertz's law \cite{gompertz}.  Our model mortality curve, with default parameters, is shown with filled black circles.   After age 50, our model exhibits a Gompertz-like increase in mortality rate. Our model includes no developmental details, and so misses the peak of early-childhood mortality.  In the inset we show that the mortality rate monotonically increases with $F_{30}$. The increase is approximately exponential, as reported for clinical data \cite{Mitnitski2006}.

In Fig.~\ref{f:FIdata} we show the frailty index vs age, as determined from the Canadian National Public Health Survey (NPHS) from 1994-2011 \cite{Mitnitski2013} (red circles with red line). Notable is a significant upward curvature after age 60. Also shown are model $F_n$, for $n=30$ or $799$, (black or small green circles, respectively) with similar upward curvature at later ages, due to our non-zero interaction $\sigma$. Our model starts at $t=0$ with all $d_i=0$, so that $F=0$. We see that $F_n$ is slightly above the population $F$ data, though the agreement becomes better at smaller $n$.  The choice of the number of deficits $n$ included in the diagnostic frailty $F_n$ has  no influence on the model dynamics. However, increasing $n$ includes more nodes with lower connectivity -- and simultaneously decreases the weight given to highly connected nodes. The dependence of $F_n$ on $n$ therefore indicates that highly connected nodes are somewhat protected from accelerated damage due to interactions. 

The deficit nodes included in $F_n$ exclude the mortality node. The coloured points of Fig.~\ref{f:Gompertzdata}, as indicated by the legend, shows the model mortality rates vs. age for individuals grouped by quartiles of the frailty index. At every age, mortality is strongly affected by $F$ -- as also shown in the inset. Conversely, at a given $F$ the mortality rates are only relatively weakly dependent on age. Much (though not all) of the weak age dependence within the quartiles results from variations of $F$ within the quartile.  At younger ages ($\lesssim 60$ years), the mortality rates of the lowest quartile are nearly identical to the population average -- reflecting the low frailties at those ages. 

\begin{figure}
\centering
\begin{minipage}{.45\textwidth}
\includegraphics[trim = 15mm 10mm 5mm 10mm, clip, width=\textwidth]{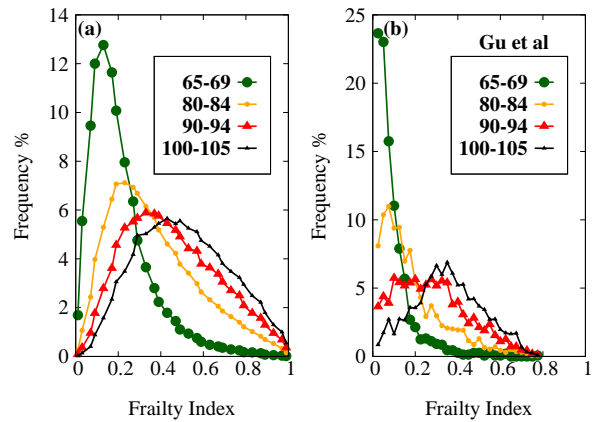}
\caption{(color online) Distribution $P(F)$ for different age ranges, as indicated in the legend. (a) model results, using $F_{30}$, (b) Chinese population data from Gu {\em et al} \cite{Gu2009}.}
\label{f:oldyoung}
\end{minipage} 
\end{figure}

We show the distribution $P(F)$ for different ages in more detail in Fig.~\ref{f:oldyoung}. We can see that both the average and peak $F$ increases with age. The shape of the distribution $P(F)$ also changes with age, evolving from a right-skewed distribution at young ages towards a more symmetrical distribution for older sub-populations \cite{Rockwood2004, Gu2009}.  On the left, in Fig.~\ref{f:oldyoung} (a), we find this pattern with our model data. The distribution does not become left-skewed, even though we must have $F \leq 1$, since mortality typically removes individuals as they reach higher $F$.  On the right, in Fig.~\ref{f:oldyoung} (b), we show Chinese population data for the same age ranges \cite{Gu2009}. While the qualitative patterns are the same, the Chinese data population data is shifted towards lower frailties. Notably, the model exhibits all values of $F \in [0,1]$ while the population data shows an apparent maximal value of $F_{max} \approx 0.7-0.8$.  Other population studies have also reported a maximal $F_{max} \approx 0.7-0.8$ \cite{maximum}. 

\begin{figure}  
\centering\begin{minipage}{.45\textwidth}
\includegraphics[trim = 19mm 15mm 10mm 15mm, clip, width=\textwidth]{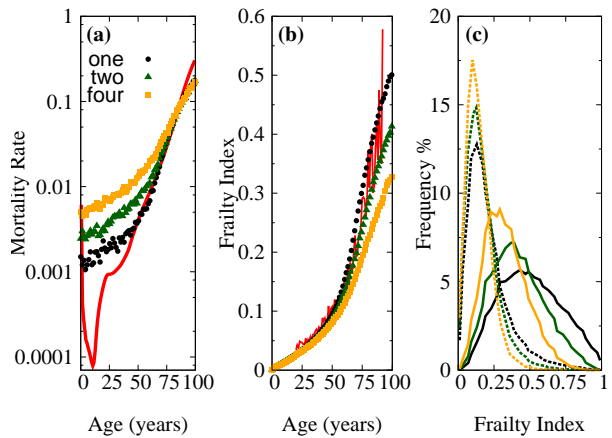}
\caption{(color online) Model results with modified mortality that occurs when {\em any} of the top one, two, or four nodes are damaged (black circles, green triangles, and yellow squares respectively). The parameters are otherwise the same as our default parameterization. Red lines correspond to population data. (a) Mortality rate vs age, (b) $F_{30}$ vs age, (c) model data frailty distribution, $P(F_{30})$ vs $F_{30}$, for age cohorts between 65-69 years (dashed lines) and between 100-105 years (solid lines).}
\label{f:OR}
\end{minipage}
\end{figure}

Fig.~\ref{f:OR} shows the results of having mortality result from the damage of {\em any} of the top $1$, $2$, or $4$ connected nodes. Note that using one node is our standard mortality rule used elsewhere in the paper. The peak of the death-age distribution is rescaled to $87.5$ years in each case, and this  leads to continued agreement of mortality rates at older ages. With this logical-``OR'' with multiple mortality nodes, we see in (a) that the mortality at younger ages increases. However, in (b) the agreement with $F$ vs age is improved. We also see in (c) that an approximate $F_{max} \approx 0.7-0.8$ emerges when the top $4$ nodes are used for mortality. Essentially, it becomes increasingly unlikely to have more than one of the top (mortality) nodes undamaged and yet to still have large $F$. We note that in our model there is no strict frailty maximum, and so multiple-node ``OR'' may not be the underlying cause of the population frailty maximum. Clearly the mortality condition is important, and is a sensitive control of qualitative model behavior. [More data with a multi-node logical-``OR'' mortality is shown in the appendix in Fig.~\ref{OR} while the complementary multi-node logical ``AND'' mortality is explored in Fig.~\ref{AND}.]  

\begin{figure} 
\centering\begin{minipage}{.45\textwidth}
\includegraphics[trim = 15mm 10mm 5mm 10mm, clip, width=\textwidth]{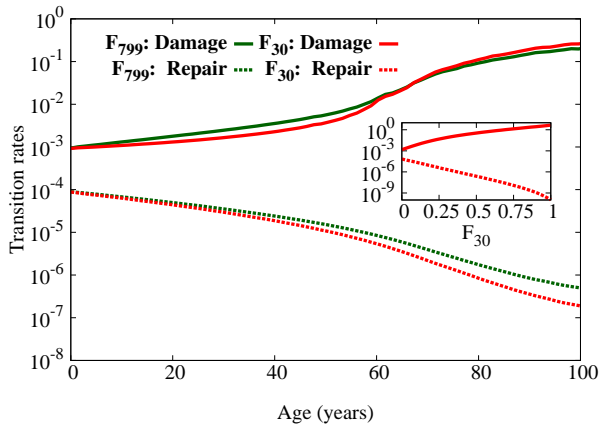}
\caption{(color online) Damage ($\Gamma_+$) and repair ($\Gamma_-$) rates vs age, averaged over undamaged or damaged frailty nodes, respectively,  corresponding to $F_{30}$ or $F_{799}$ as indicated.  All rates are per year. The inset shows the same rates vs $F_{30}$.}
\label{f:rates}
\end{minipage}
\end{figure}

Fig.~\ref{f:rates} shows our model damage ($\Gamma_+$) and repair ($\Gamma_-$) rates vs. age, where rates are per year. Increasing age monotonically increases the damage rate, and decreases the repair rate. To obtain $\Gamma_\pm$ vs age, we have averaged the local rates over the indicated frailty nodes for the surviving population. The inset shows the same rates vs. $F_{30}$. They are consistent with the similar trends in $\Gamma_\pm$ vs local frailty $f_i$ shown in Fig.~\ref{f:DoubleWell} (c), together with the average increase of $F$ with age.

\section{Discussion} 
The frailty index $F$ is a quantitative measure of individual aging that is as informative as chronological age in predicting health outcomes, including mortality. The FI is defined as the fraction of damaged health-related deficits.  Given the increasing health burden on our burgeoning elderly population, it is important to better understand the observed quantitative dynamics of the frailty index. To do this, we have developed a stochastic  model of damage and repair of interacting deficits. The stochasticity of $F$ is  important, and we have qualitatively recovered the observed broadening of the distribution $P(F)$ with age (see Fig.~\ref{f:oldyoung}). Stochasticity can, in principle, arise from either stochastic damage or stochastic repair -- or both. We have  found that  a significant amount of explicit deficit repair is consistent with observed age-dependent mortality and $F$ (see appendix Fig.~\ref{R}). 

We have addressed our two key questions. First, can the observed upward curvature of $F$ vs age be obtained with time-independent interactions between deficits, or must explicitly time-dependent damage rates be invoked?  We have shown (see Fig.~\ref{f:FIdata}) that time-independent interactions ($\tilde{\sigma}$) alone are sufficient. Second, can we recover the observed connection between $F$ and mortality without building $F$ directly into model mortality? We have shown that this is indeed possible (see Fig.~\ref{f:Gompertzdata}), since the (highly connected) deficits we have used for mortality have no overlap with the (less connected) deficits that we have used for our frailty measure $F_n$. 

Due to deficit interactions, our model repair rates are age dependent (see Fig.~\ref{f:rates}).   While strong age-dependent slowing in human wound-healing rates have been reported since the seminal work of Lecomte du No\"{u}y \cite{duNouy1916}, more recent emphasis has been on the proximal mechanisms controlling these rates -- resulting in questions about how or whether to attribute slowed healing to age per se \cite{healingreview}.  Nevertheless,  age-related slowing of wound-healing is also reported in e.g. mice \cite{Yanai2015}. Echoing this, bone fracture repair is impaired in older rats \cite{Meyer2001}, though recent emphasis has been on detailed mechanisms behind this slowing \cite{Gruber2006}. Indeed, there are many age-related changes that could be considered even at the cellular level \cite{LopezOtin2013}.  Our model provides a coarse-grained picture of  interactions between deficits, and indicates that age-related slowing may be driven by frailty-related slowing. Given the large variability of $F$ at a given age (see Fig.~\ref{f:oldyoung}), we conclude that $F$ is important to control for when assessing age-dependent repair rates. 

It is attractive to think that highly connected model deficits correspond to clinically-accessible high-level physiological states. Indeed, we use our most connected node to indicate mortality and reserve the next-most connected nodes for $F_n$. However, there is no direct equivalence between model deficits and any specific physiological deficits. Indeed, we may best think of our model deficits as  combinations of physiological measures. Similarly, we might think of conveniently accessible clinical deficits (see e.g. \cite{Mitnitski2001}) as reflecting combinations of cellular, organ, and systems-level mechanisms (see e.g. \cite{biochem}). If so, the effective connectivity of individual clinical deficits may be difficult to directly assess. 

In the appendix, we have systematically varied each of our key model parameters to see how they affect mortality, $F$ and deficit transition rates. The default parameters used in our paper were found to give reasonable qualitative agreement with current data \cite{sloppy}. We have illustrated how including more highly connected nodes in our mortality rules can improve $F$ evolution and distribution (see Fig.~\ref{f:OR}) -- albeit at the expense of further overestimating mortality rates at younger ages.  This was done by triggering mortality if {\em any} of multiple mortality nodes are damaged. Complementing this, in Fig.~\ref{AND} we have explored triggering mortality only if {\em all} of multiple mortality nodes are damaged. In that case, we can get better agreement with mortality rates but worse agreement with $F$ behavior. The sensitivity of our model results to mortality rules is promising since it will allow us to better explore the connection between model deficits and mortality and so between clinical deficits and human mortality. 

While we have found that both the average and the distribution of connectivities affect our model behavior when other parameters are kept fixed (see Fig.~\ref{K} and \ref{NETWORK}), we caution that we have not exhaustively searched parameter space and so cannot rule out alternative topologies \cite{sloppy}. We have also not explored  differences between directed (see e.g. \cite{Vural2014}) and non-directed network links. In this work we have only used non-directed links, and have predominately considered scale-free networks. In future work, we will use our model to develop new  diagnostics that are more sensitive to network topologies and apply them to clinical data in order to better characterize effective interaction networks of clinical deficits. 

The emerging prospects of large quantities of laboratory or biomarker data \cite{biochem}, as well as tracked health data \cite{bigdata}, usable for assessing frailty present the question of how to rationally incorporate new data when it is available to best contribute towards health assessment. In future work, we will use our model to learn how to construct a better $F$ that is more predictive of individual mortality. Such a frailty measure may include aspects of individual frailty trajectories, but also an optimized compromise between number and quality of deficits. Our computational model will allow us to start this development with high-quality model data, before we test our insights against current and emerging clinical data.  

\begin{acknowledgements}
We thank the Atlantic Computational Excellence Network (ACENET) and Westgrid for computational resources. ADR thanks the Natural Sciences and Engineering Research Council (NSERC) for operating grant RGPIN-2014-06245, ABM thanks Canadian Institutes of Health Research (CIHR) for grant MOP24388, and both ABM and KR thank the Capital Health Research Fund for support.  We thank Dr. Danan Gu for sending us population data for Fig.~\ref{f:oldyoung} \cite{Gu2009}.
\end{acknowledgements}

\appendix*   
\section{Parameter scan}

In this appendix we systematically explore the effects of varying our model parameters.   We show the effects of varying parameters around our default parameter values or mortality rules. For each figure, we show (a) mortality rate vs age for indicated parameter values, with the solid red line indicating 2010 U.S. population data \cite{USAdata}, (b) $F$ vs age for indicated parameter values, where the red circles indicate Canadian population data \cite{Mitnitski2013}, (c) frailty distributions $P(F)$ at ages $40-49$ years and $90-99$ years for indicated parameter values, and (d) transition rates $\Gamma_\pm$ vs age for indicated parameter values. 

\begin{figure}[!htbp]
  \begin{center}
      \resizebox{70mm}{!}{\includegraphics[trim = 15mm 10mm 5mm 20mm, clip, width=\textwidth]{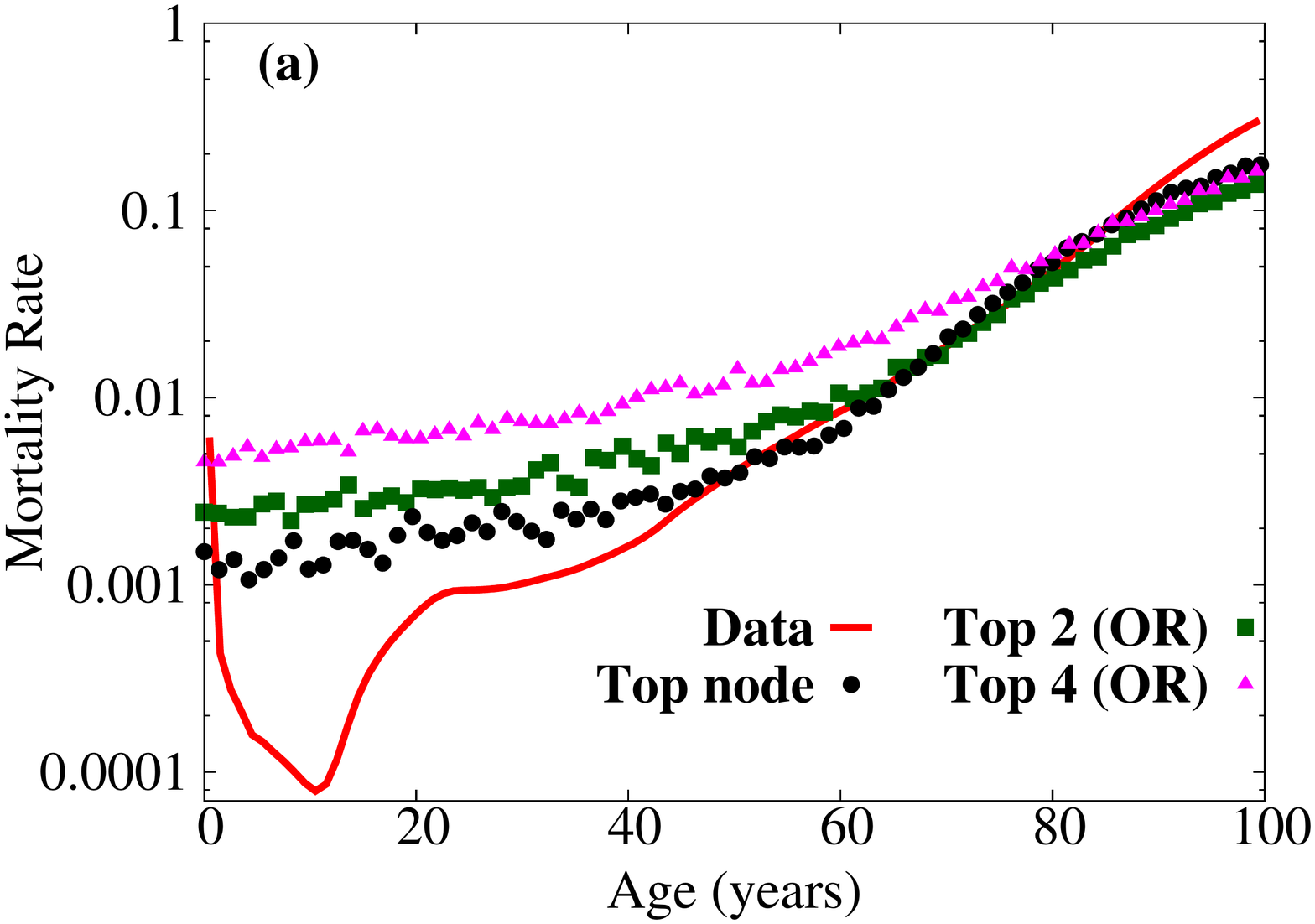}} \\
      \resizebox{70mm}{!}{\includegraphics[trim = 15mm 10mm 5mm 20mm, clip, width=\textwidth]{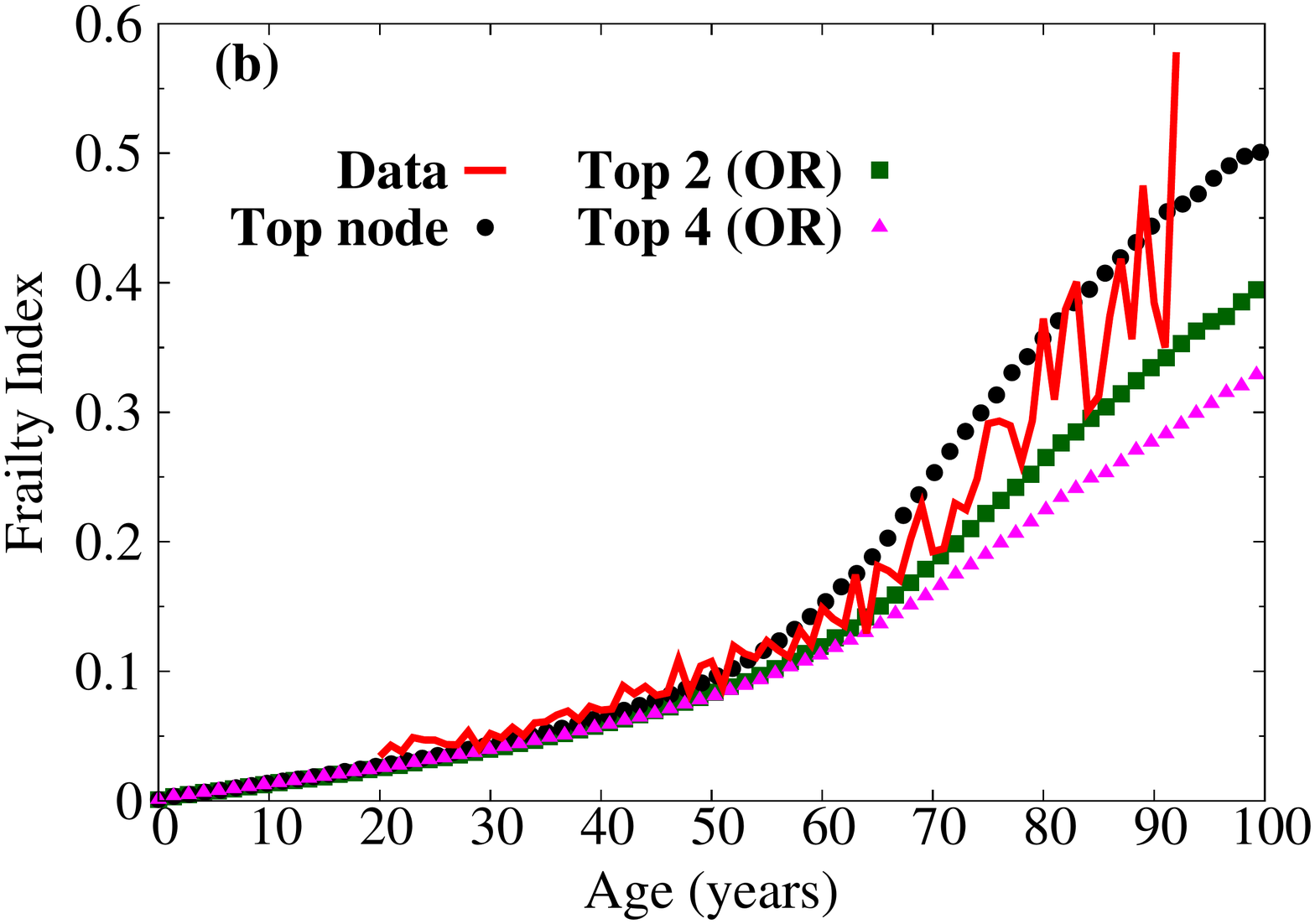}} \\
      \resizebox{70mm}{!}{\includegraphics[trim = 15mm 10mm 20mm 20mm, clip, width=\textwidth]{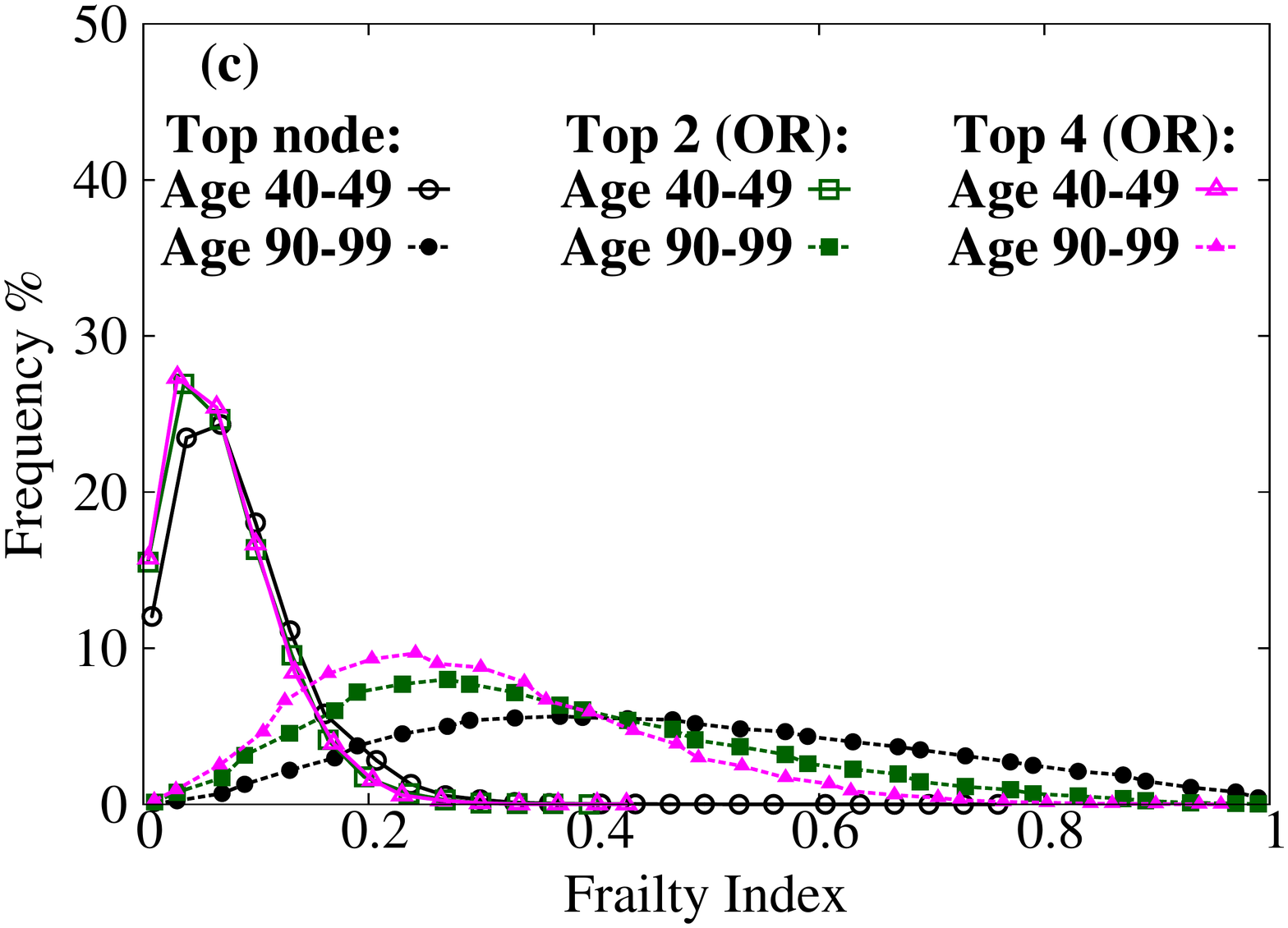}} \\
      \resizebox{70mm}{!}{\includegraphics[trim = 15mm 10mm 20mm 20mm, clip, width=\textwidth]{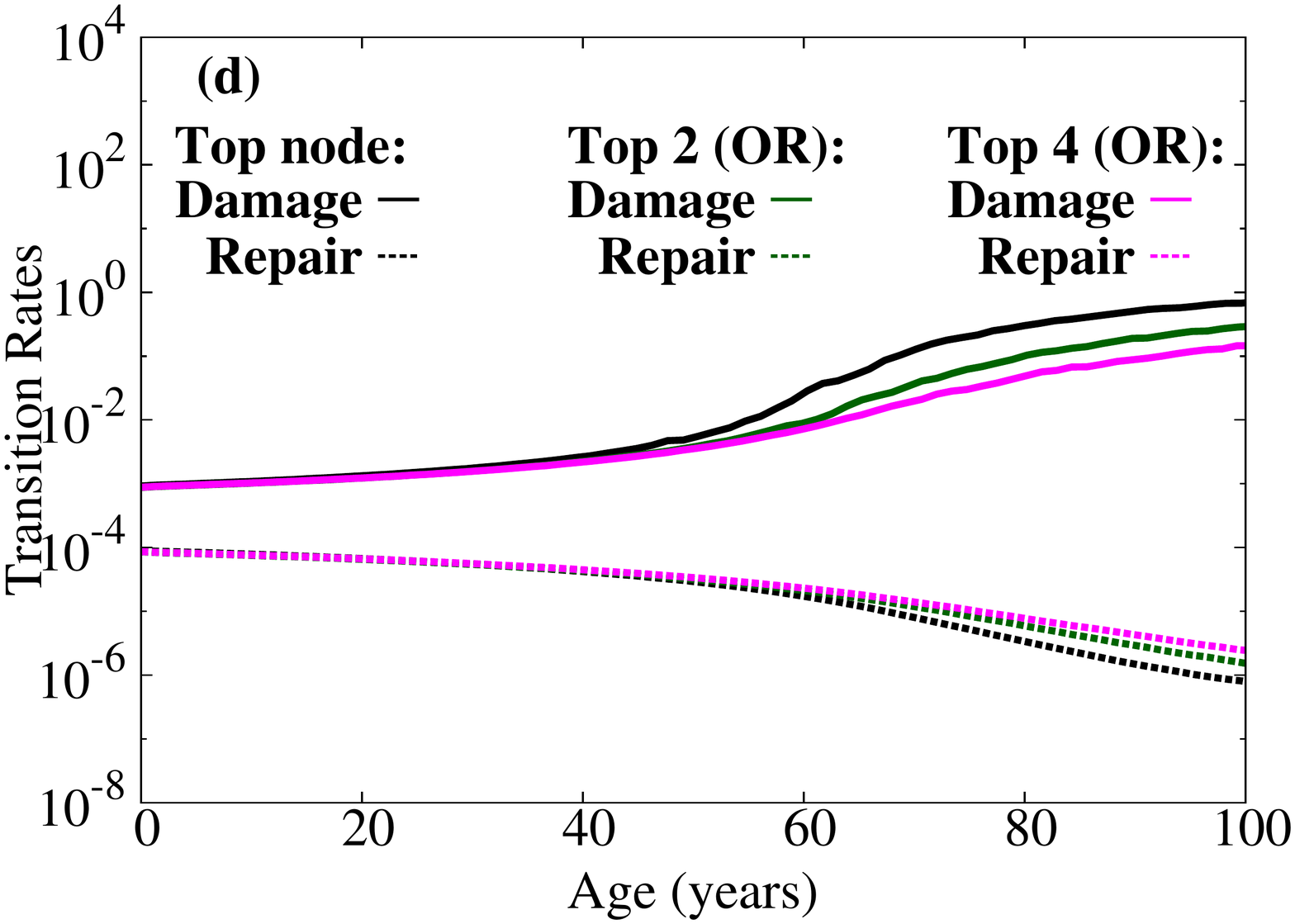}} \\
    \caption{We have explored the effect of multiple mortality nodes (one, two, or 4 nodes with black circles, green squares, or pink triangles, respectively), where mortality occurs when {\em any} of the mortality nodes are damaged (like a logical ``OR''). This was  presented in a more compressed format in Fig.~5.  One mortality node (black circles) corresponds to our default mortality rule. (a) mortality rate vs age, (b) $F$ vs age, (c) frailty index distribution $P(F)$ for young (ages $40-49$ years) and old (ages $90-99$ years) cohorts, and (d)  transition rates $\Gamma_\pm$ vs age, averaged over the $F_{30}$ nodes. We see that using more nodes in ``OR'' mortality leads to higher mortality at earlier ages, but improves $F$ vs age and also introduces a gap at larger $F$.}
    \label{OR}
  \end{center}
\end{figure}

\begin{figure}[!htbp]
  \begin{center}
      \resizebox{70mm}{!}{\includegraphics[trim = 15mm 10mm 5mm 20mm, clip, width=\textwidth]{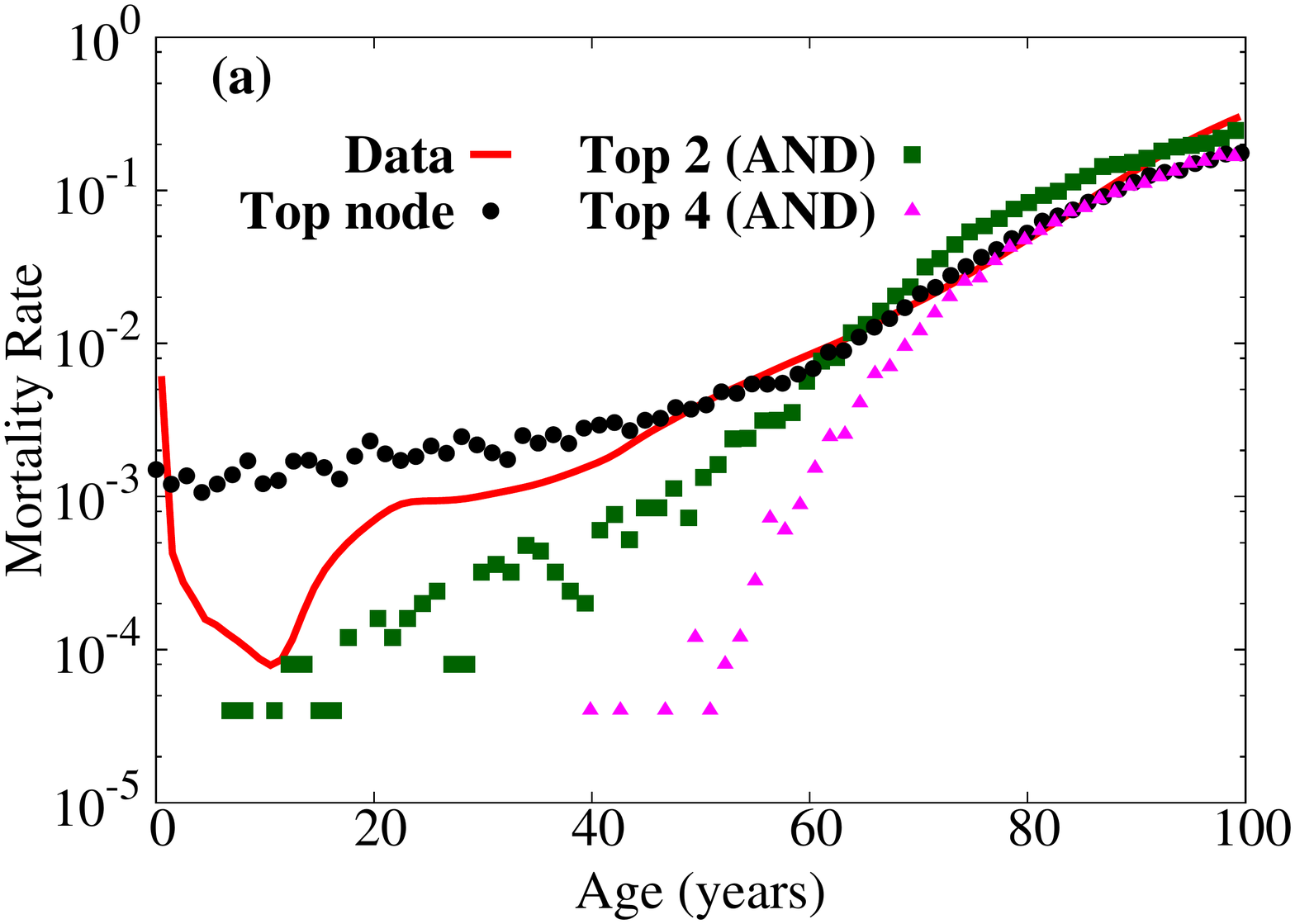}} \\
      \resizebox{70mm}{!}{\includegraphics[trim = 15mm 10mm 5mm 20mm, clip, width=\textwidth]{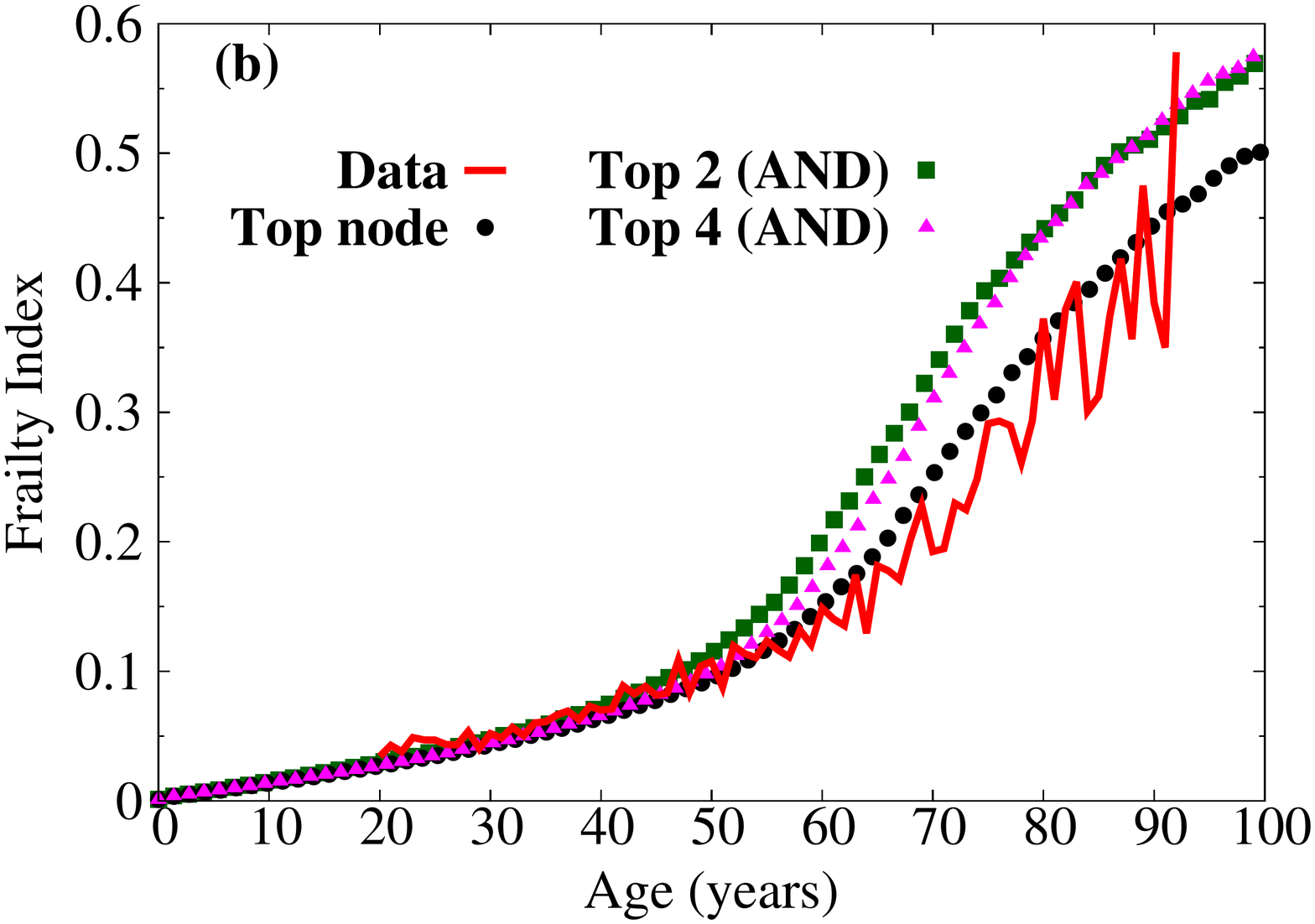}} \\
      \resizebox{70mm}{!}{\includegraphics[trim = 15mm 10mm 20mm 20mm, clip, width=\textwidth]{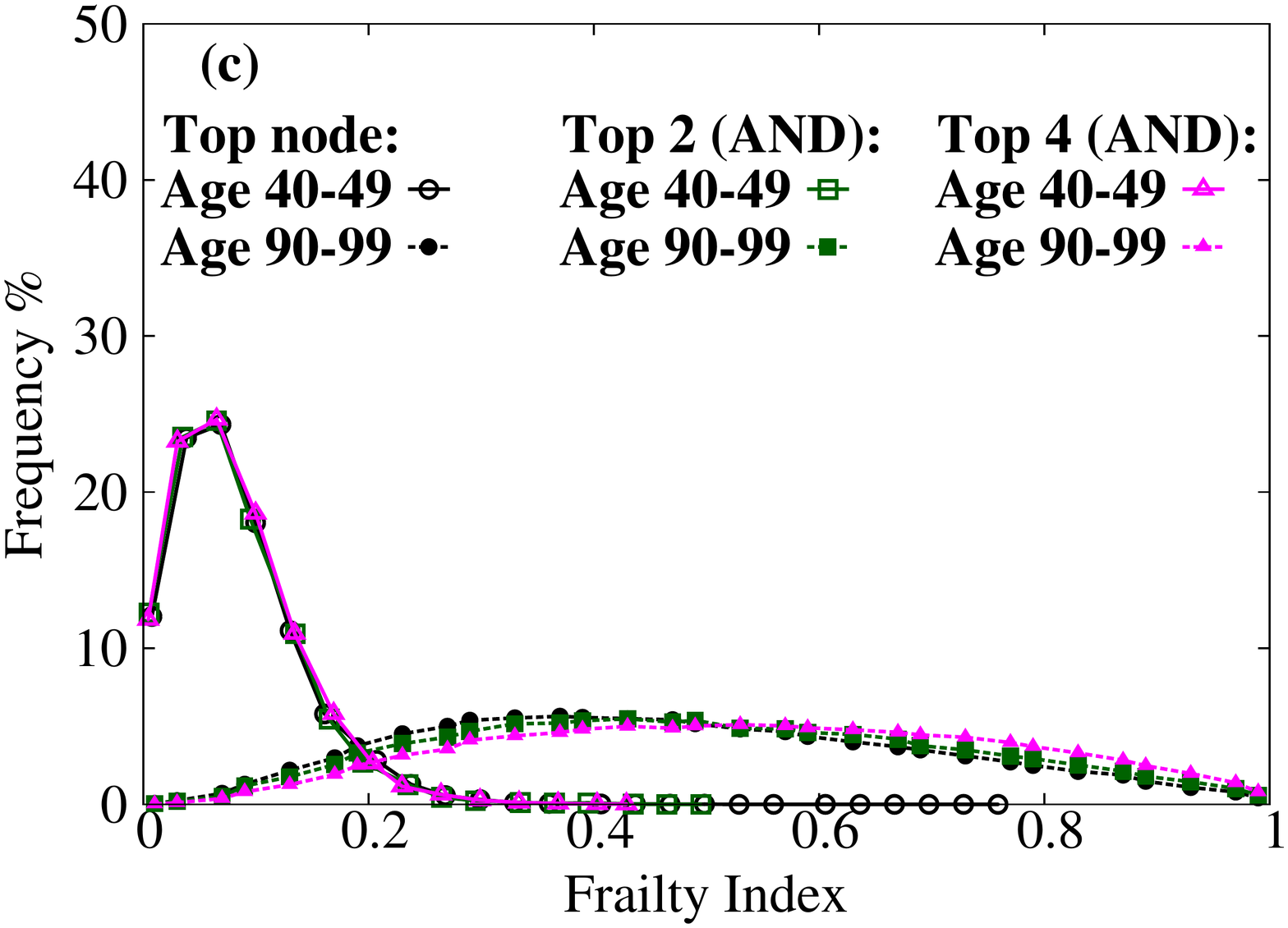}} \\
      \resizebox{70mm}{!}{\includegraphics[trim = 15mm 10mm 20mm 20mm, clip, width=\textwidth]{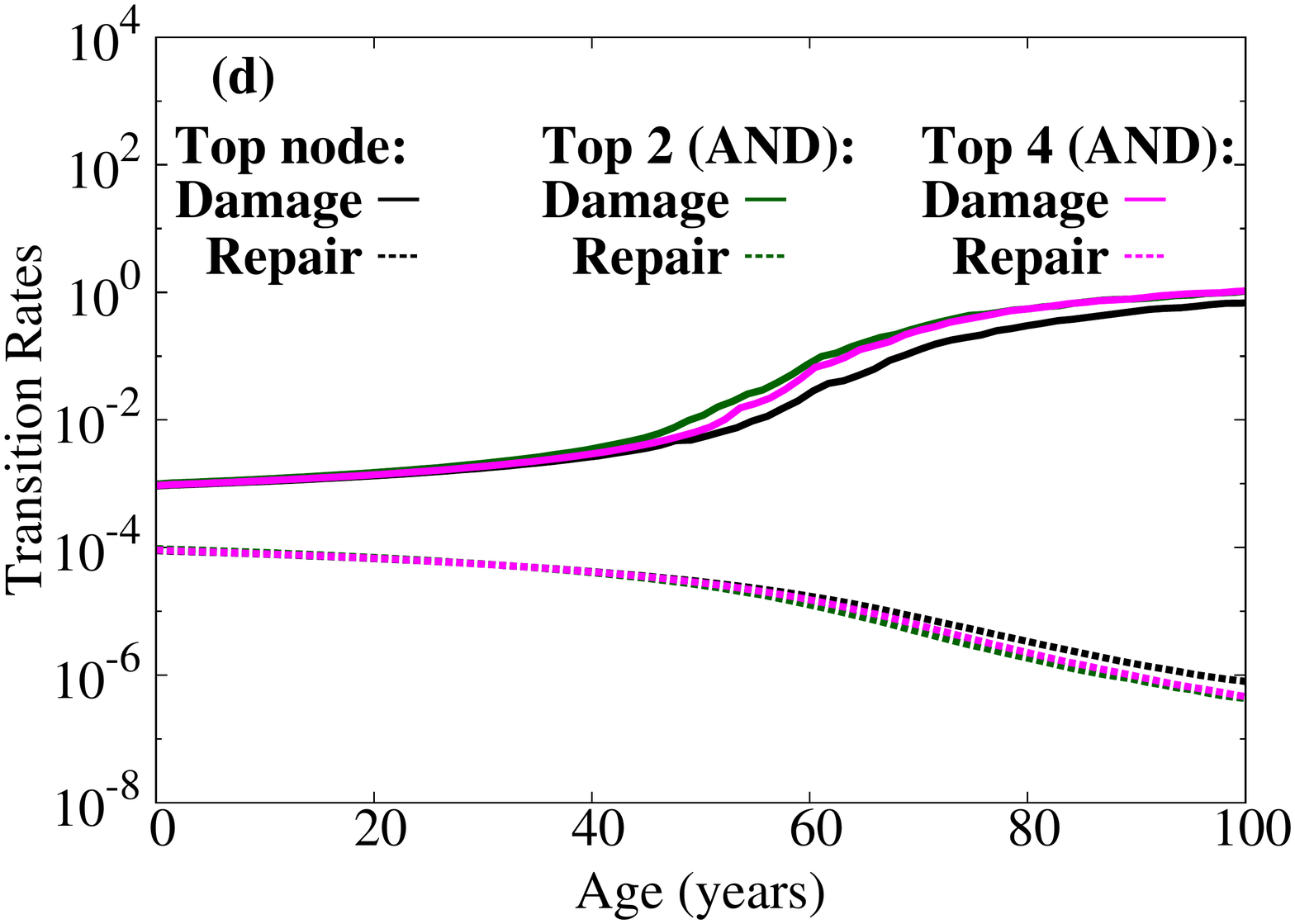}} \\
    \caption{We have explored the effect of multiple mortality nodes (one, two, or 4 nodes with black circles, green squares, or pink triangles, respectively), where mortality occurs when {\em all} of the mortality nodes are damaged (a logical ``AND'').  One mortality node (black circles) corresponds to our default mortality rule. We see that using more nodes in ``AND'' mortality leads to lower mortality at earlier ages, but at the expense of worse $F$ vs age.}
    \label{AND}
  \end{center}
\end{figure}

\begin{figure}[!htbp]
  \begin{center}
      \resizebox{70mm}{!}{\includegraphics[trim = 15mm 10mm 5mm 20mm, clip, width=\textwidth]{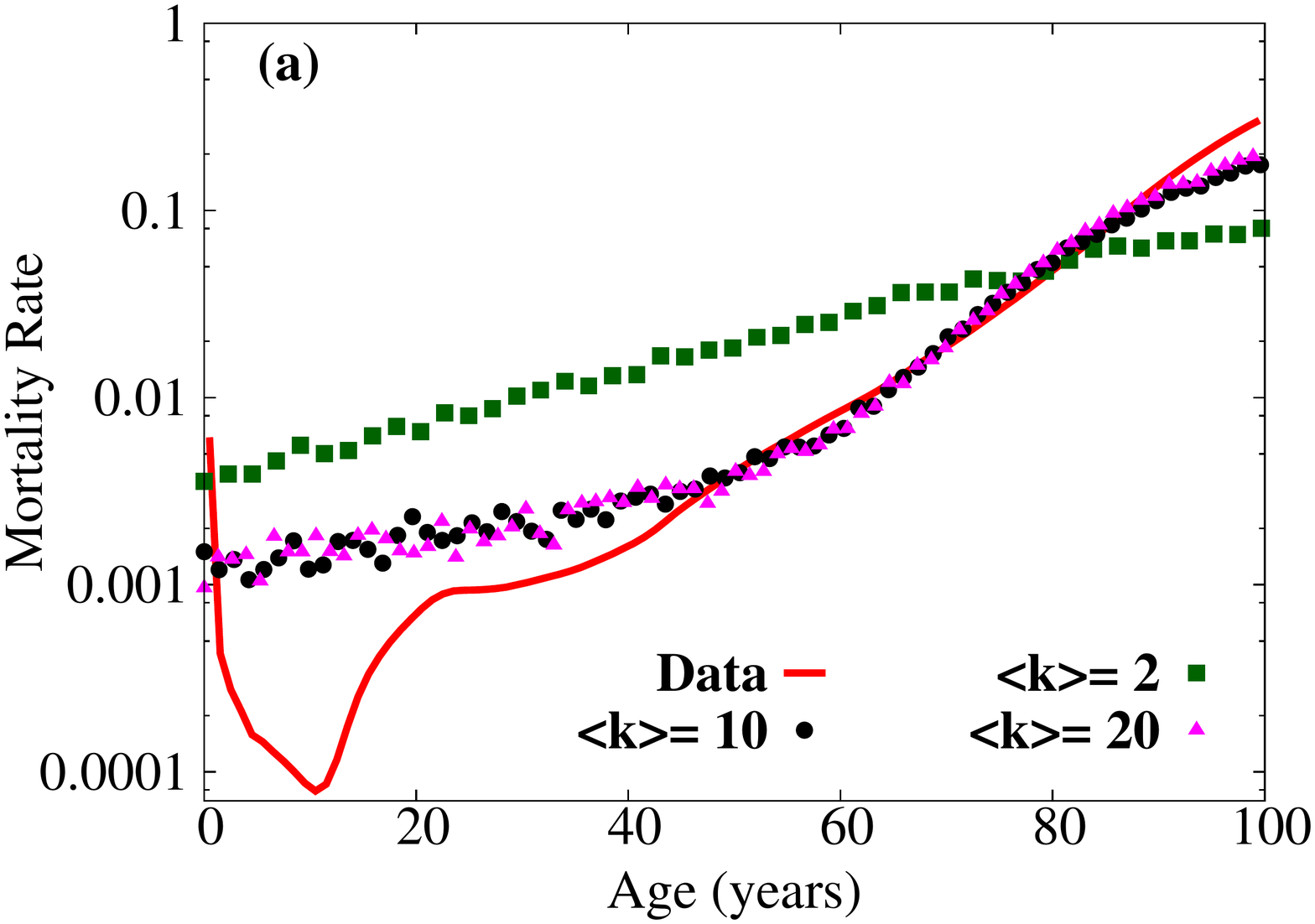}} \\
      \resizebox{70mm}{!}{\includegraphics[trim = 15mm 10mm 5mm 20mm, clip, width=\textwidth]{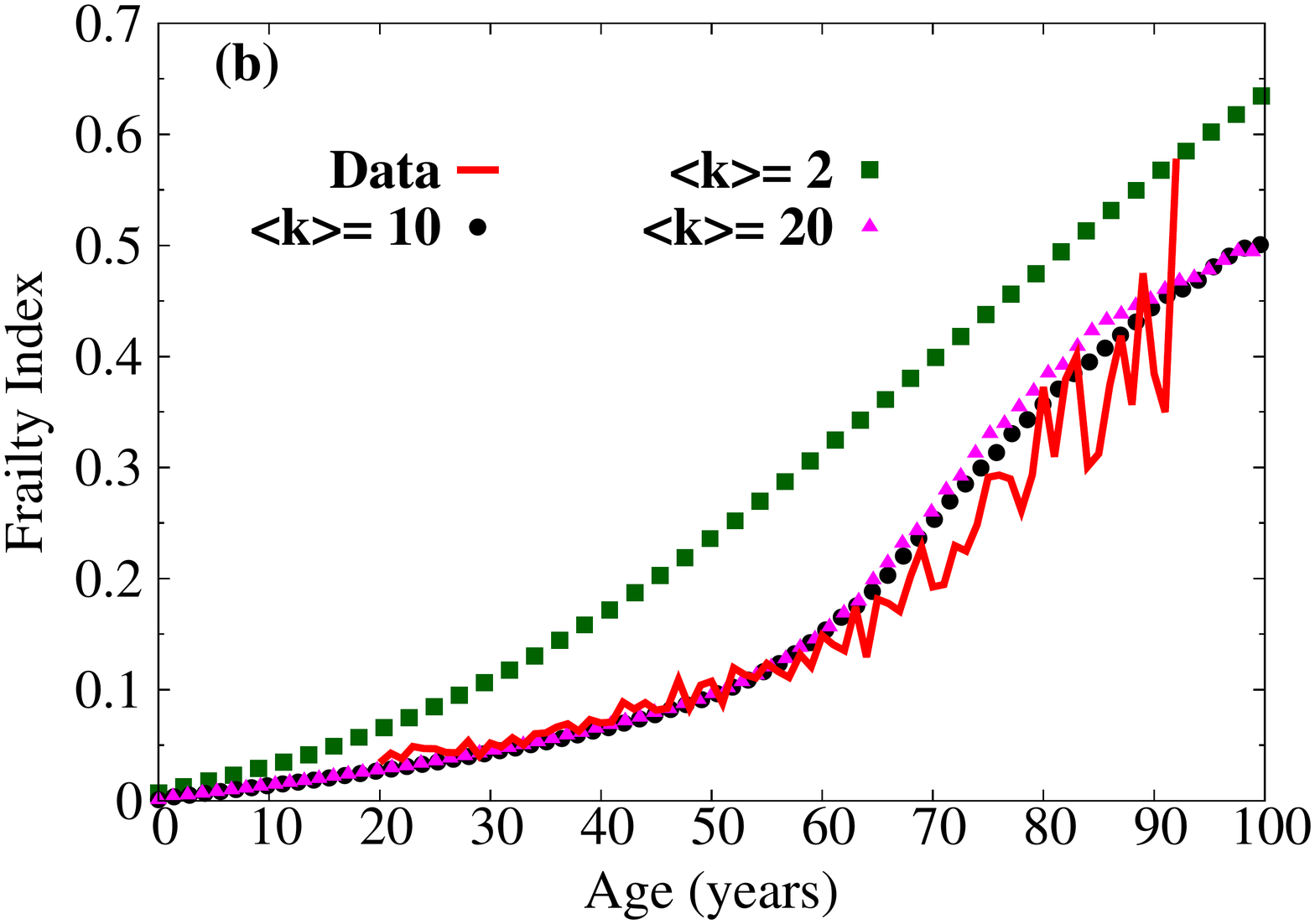}} \\
      \resizebox{70mm}{!}{\includegraphics[trim = 15mm 10mm 20mm 20mm, clip, width=\textwidth]{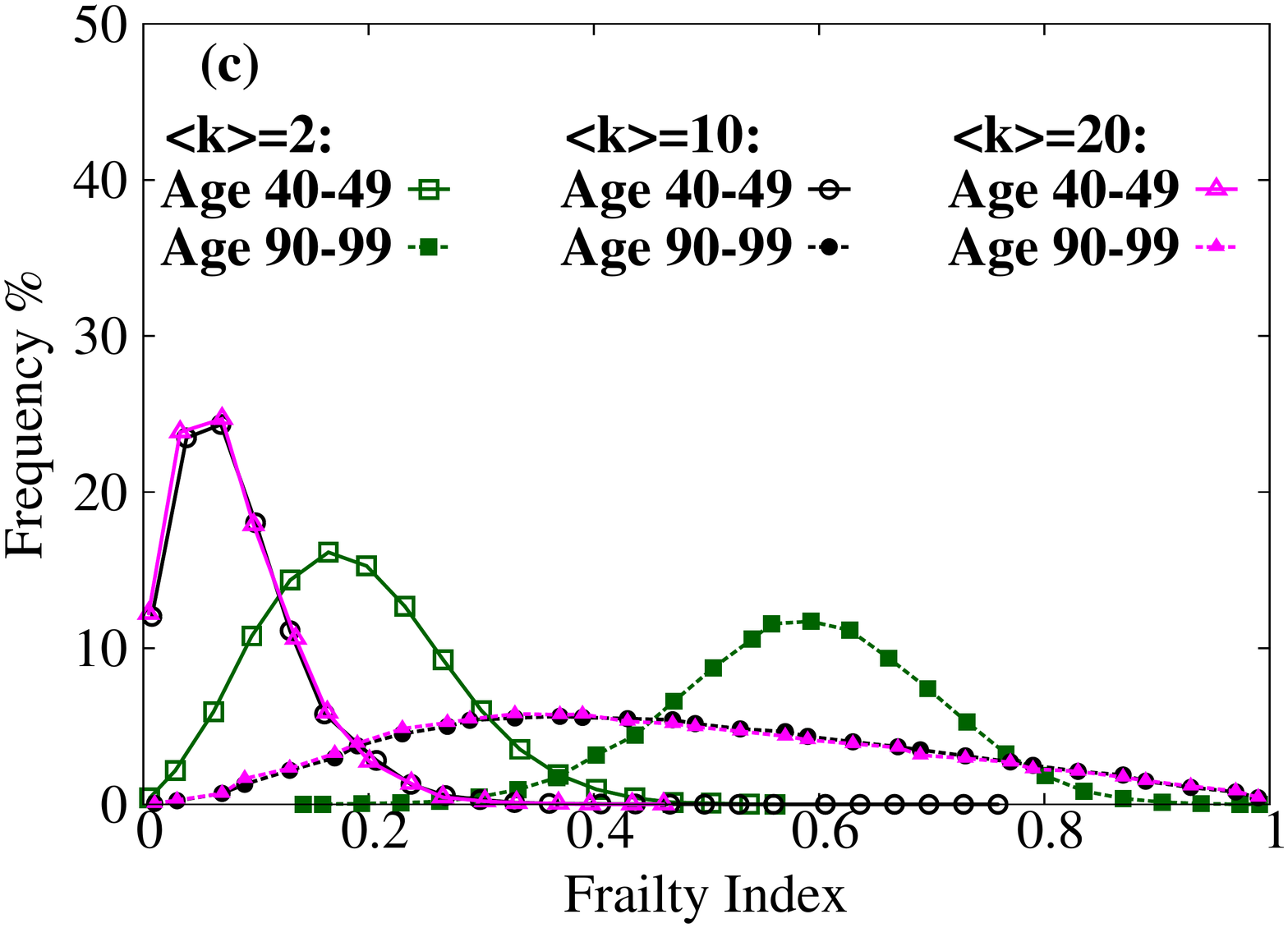}} \\
      \resizebox{70mm}{!}{\includegraphics[trim = 15mm 10mm 20mm 20mm, clip, width=\textwidth]{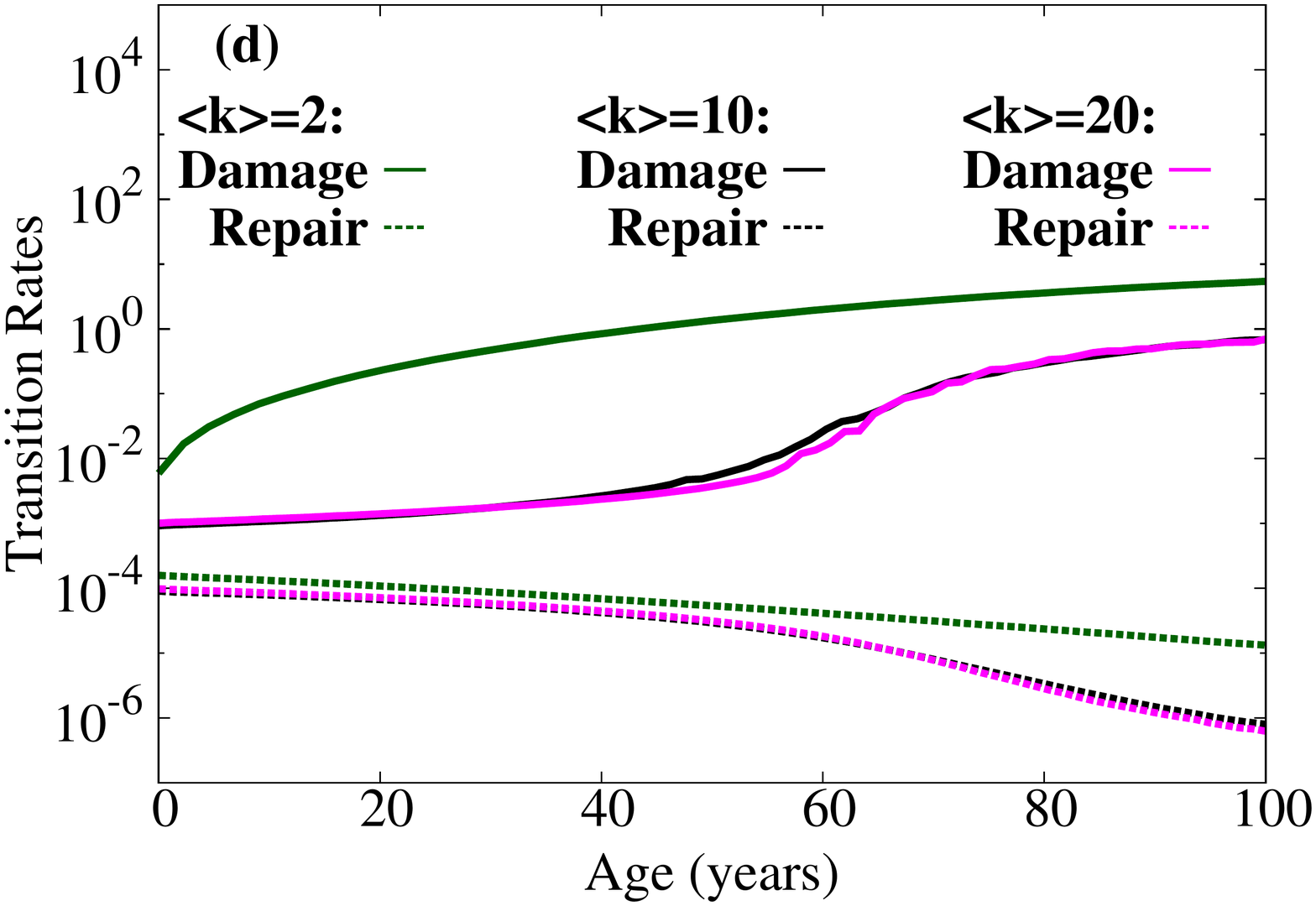}} \\
    \caption{We have varied the average degree $\langle k \rangle$ characterizing the scale-free network, with $\langle k \rangle =2$ (green squares),   $\langle k \rangle = 10$ (black circles and our default parameterization), and $\langle k \rangle=20$ (pink triangles).  (a) mortality rate vs age, (b) $F$ vs age, (c) frailty index distribution $P(F)$ for young (ages $40-49$ years) and old (ages $90-99$ years) cohorts, and (d)  transition rates $\Gamma_\pm$ vs age, averaged over the $F_{30}$ nodes.}
    \label{K}
  \end{center}
\end{figure} 

\begin{figure}[!htbp]
  \begin{center}
      \resizebox{70mm}{!}{\includegraphics[trim = 15mm 10mm 5mm 20mm, clip, width=\textwidth]{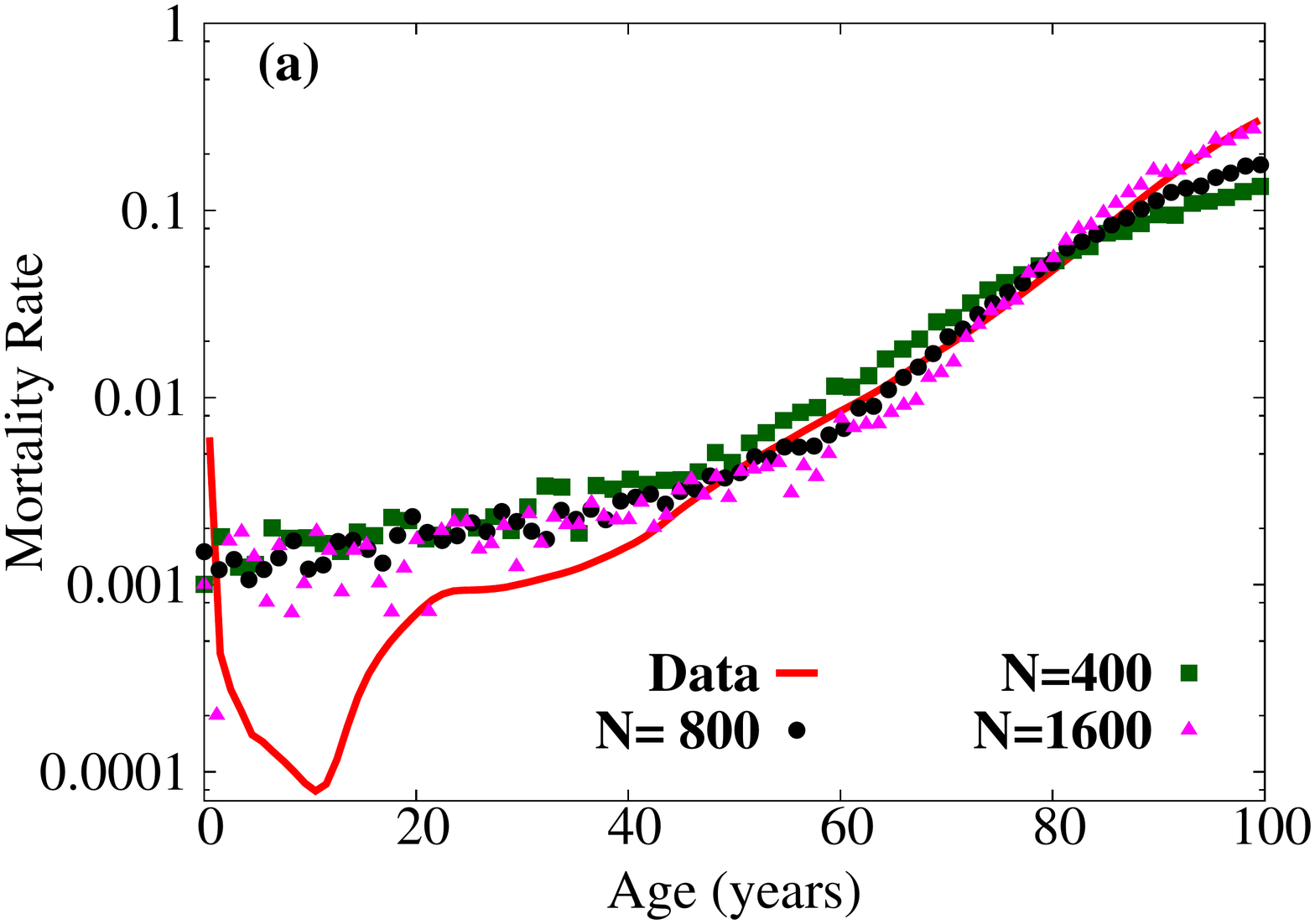}} \\
      \resizebox{70mm}{!}{\includegraphics[trim = 15mm 10mm 5mm 20mm, clip, width=\textwidth]{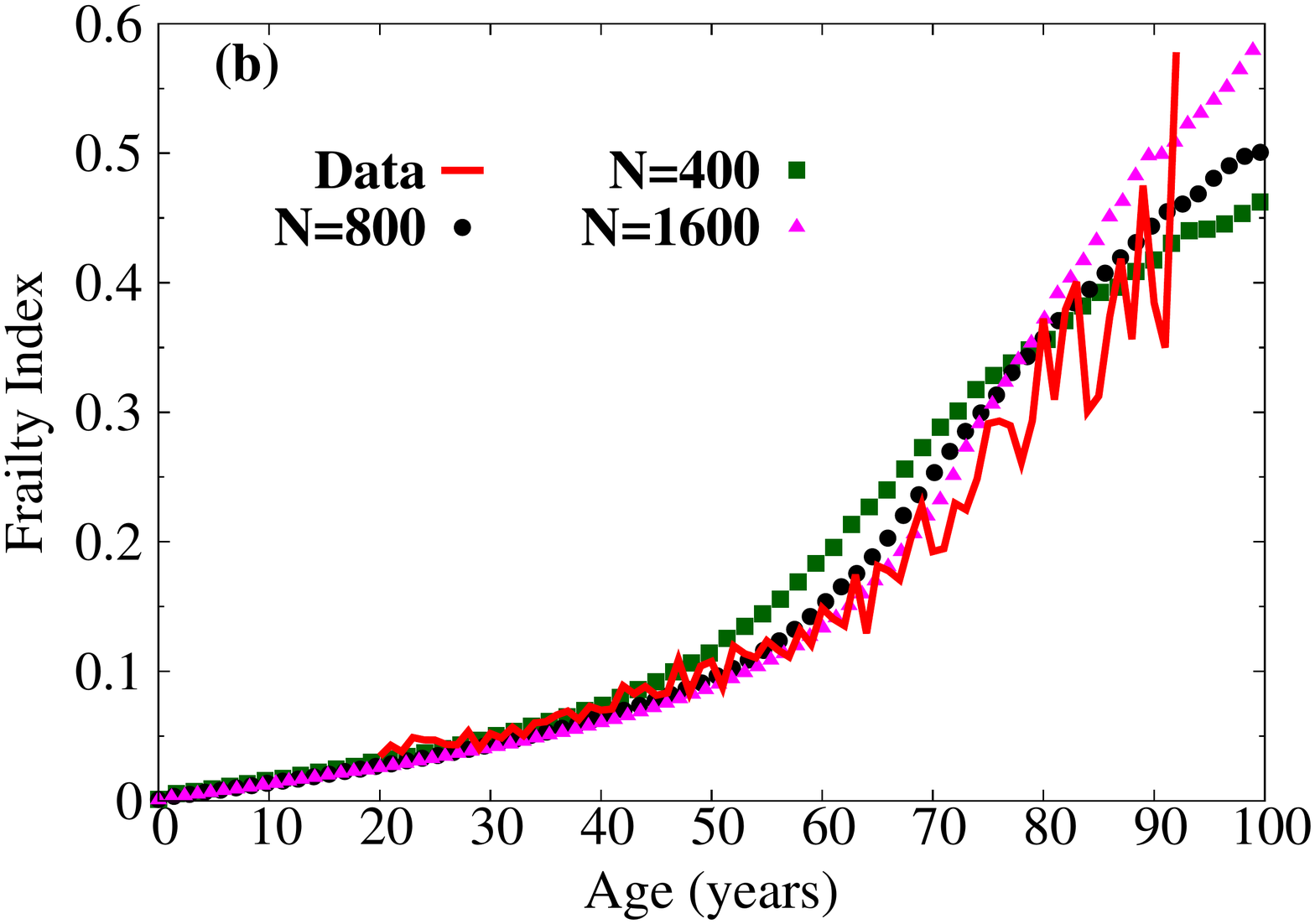}} \\
      \resizebox{70mm}{!}{\includegraphics[trim = 15mm 10mm 20mm 20mm, clip, width=\textwidth]{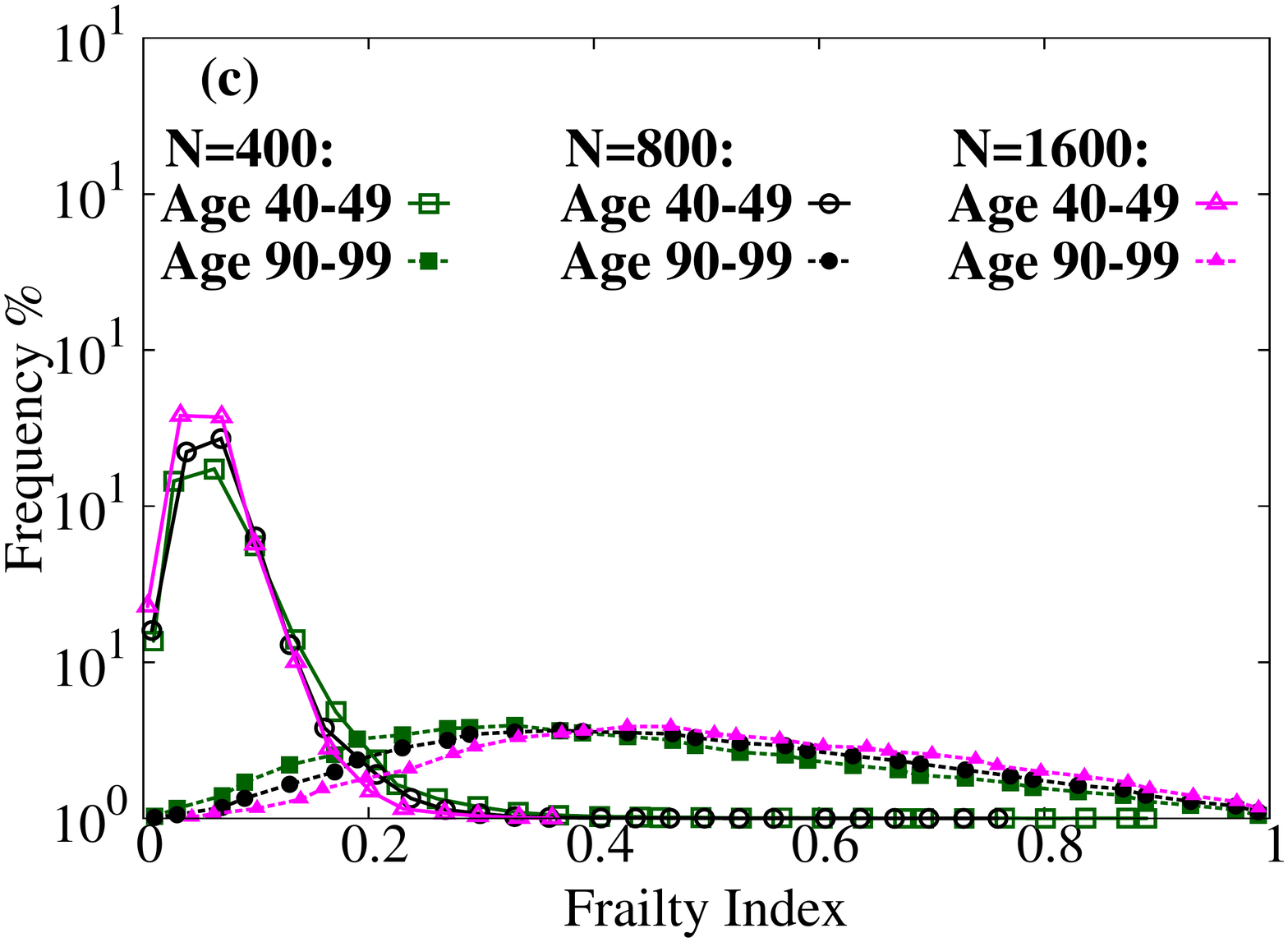}} \\
      \resizebox{70mm}{!}{\includegraphics[trim = 15mm 10mm 20mm 20mm, clip, width=\textwidth]{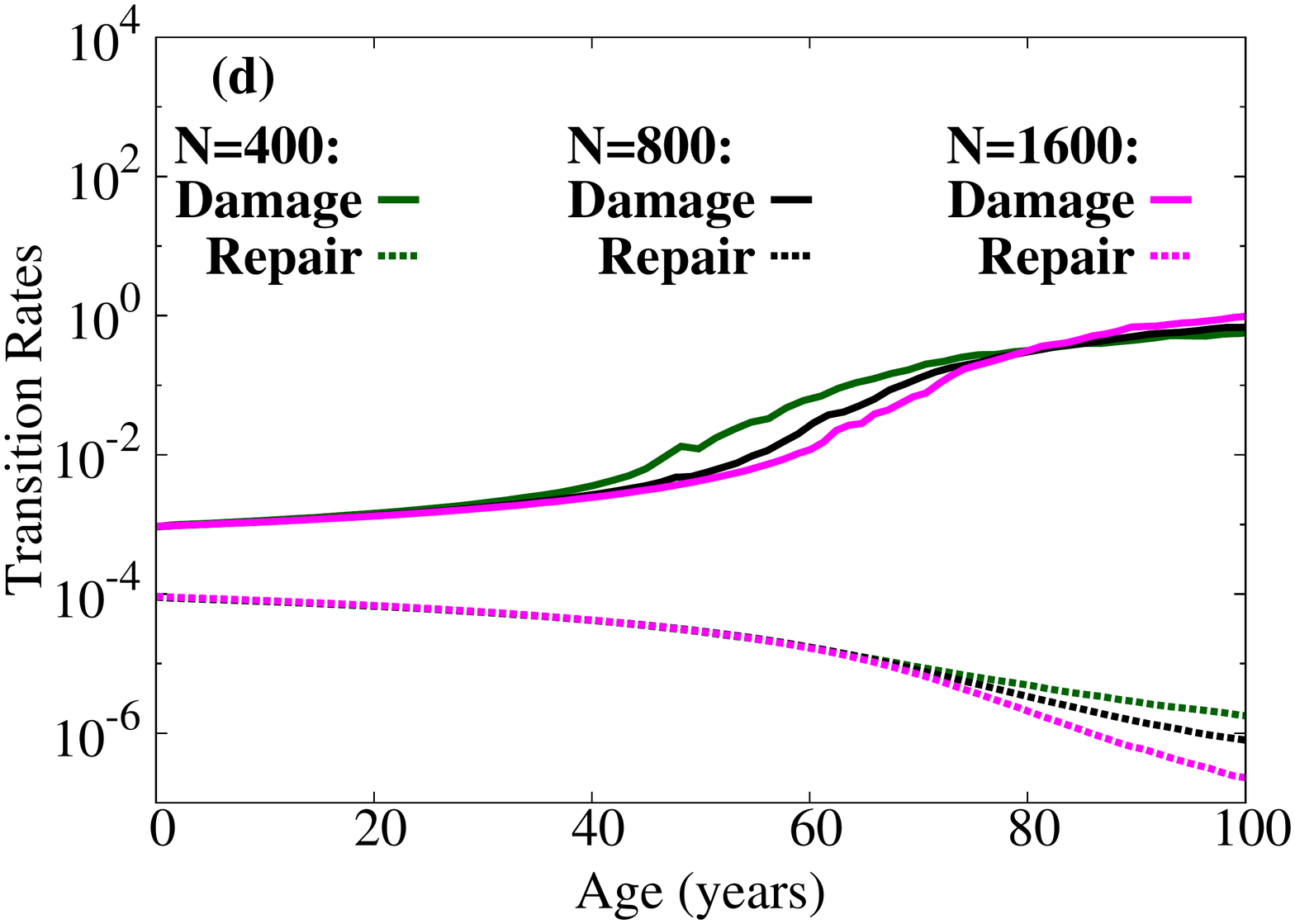}} \\
    \caption{We have varied the number of nodes $N$, with $N=800$ (black circles, our default value), $N=200$ (green squares), and $N=1600$ (pink triangles).  (a) mortality rate vs age, (b) $F$ vs age, (c) frailty index distribution $P(F)$ for young (ages $40-49$ years) and old (ages $90-99$ years) cohorts, and (d)  transition rates $\Gamma_\pm$ vs age, averaged over the $F_{30}$ nodes. We see that with sufficiently many nodes, the results are approximately independent of $N$.}
    	\label{N}
  \end{center}
\end{figure} 

\begin{figure}[!htbp]
  \begin{center}
      \resizebox{70mm}{!}{\includegraphics[trim = 15mm 10mm 5mm 20mm, clip, width=\textwidth]{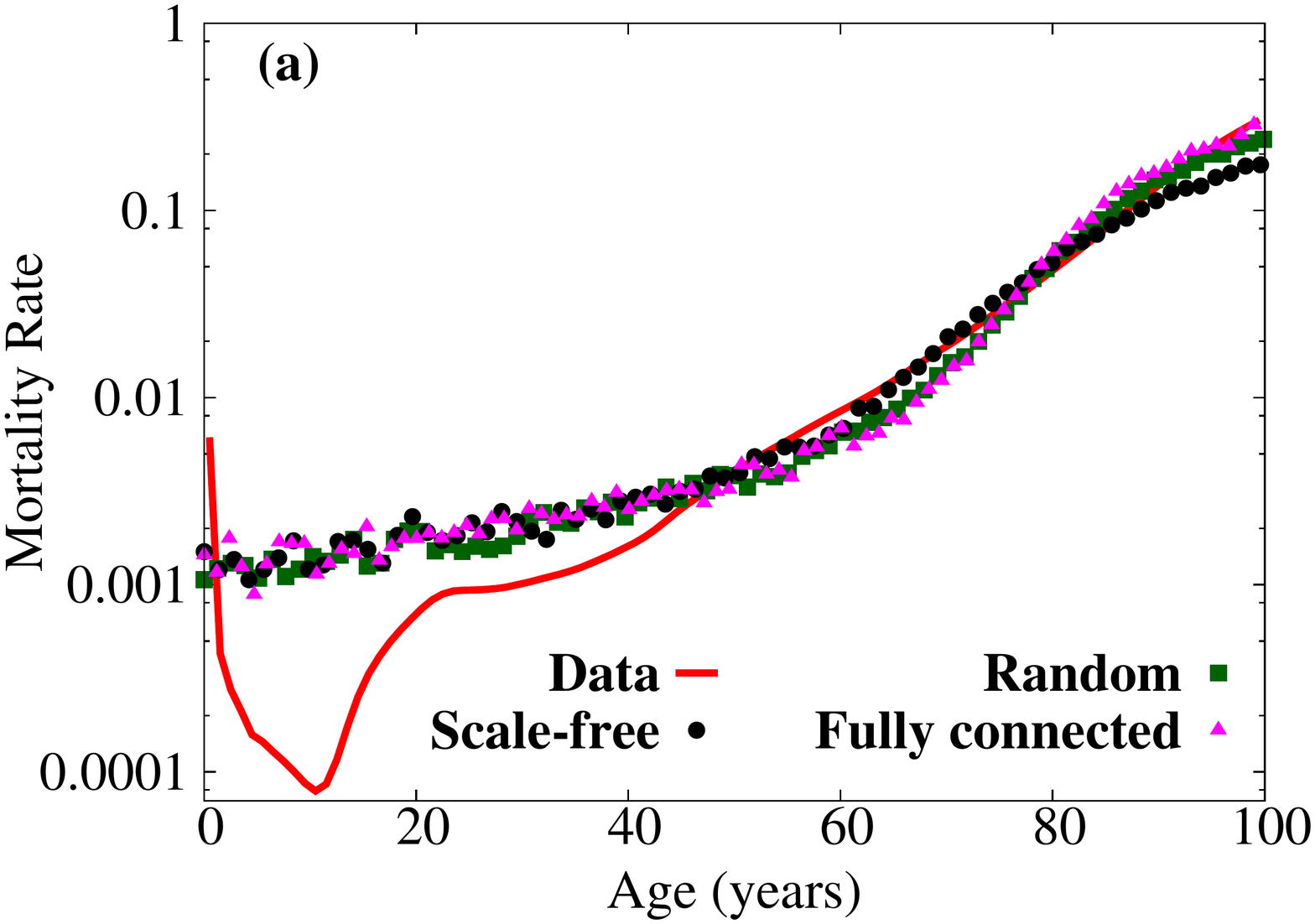}} \\
      \resizebox{70mm}{!}{\includegraphics[trim = 15mm 10mm 5mm 20mm, clip, width=\textwidth]{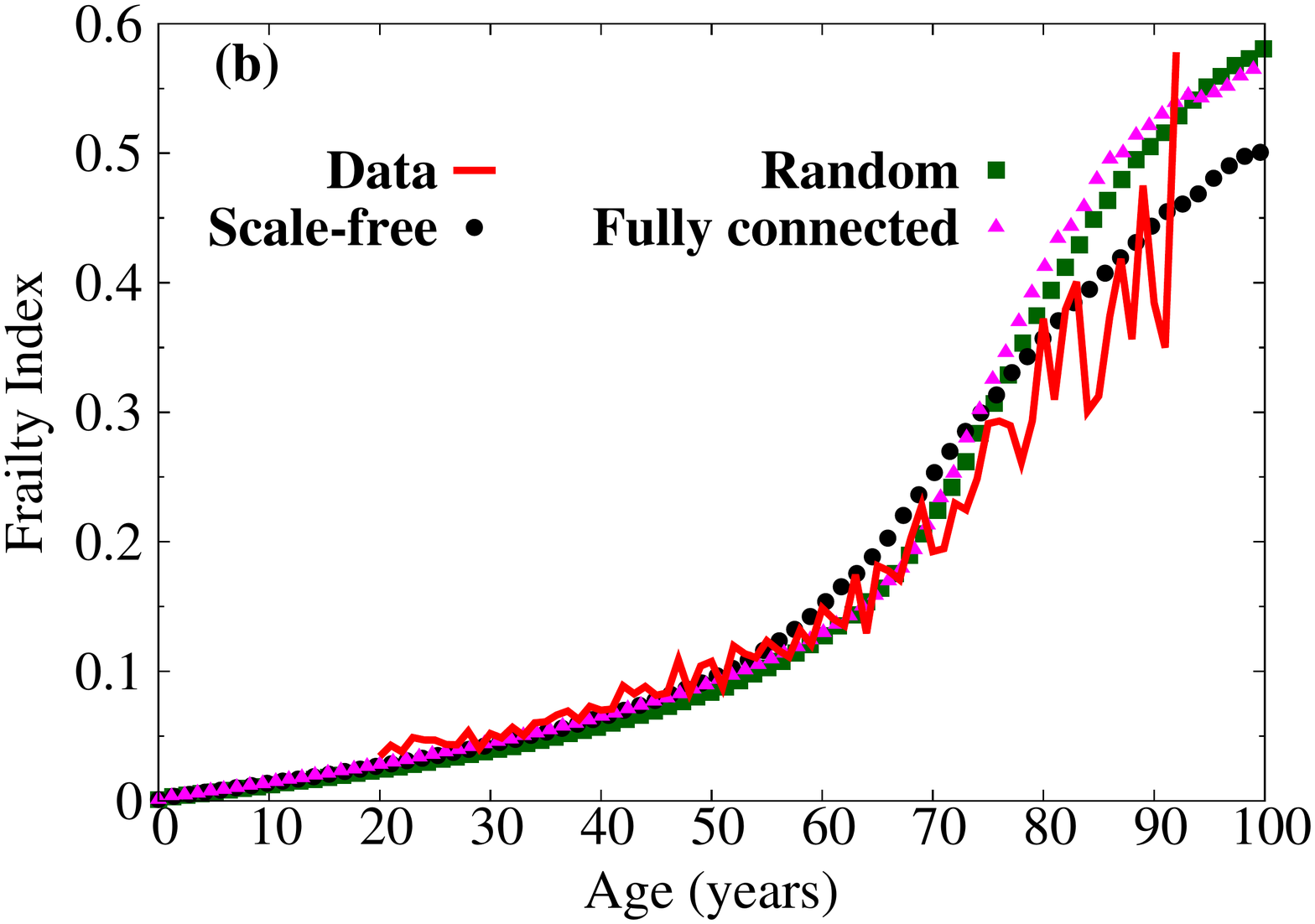}} \\
      \resizebox{70mm}{!}{\includegraphics[trim = 15mm 10mm 20mm 20mm, clip, width=\textwidth]{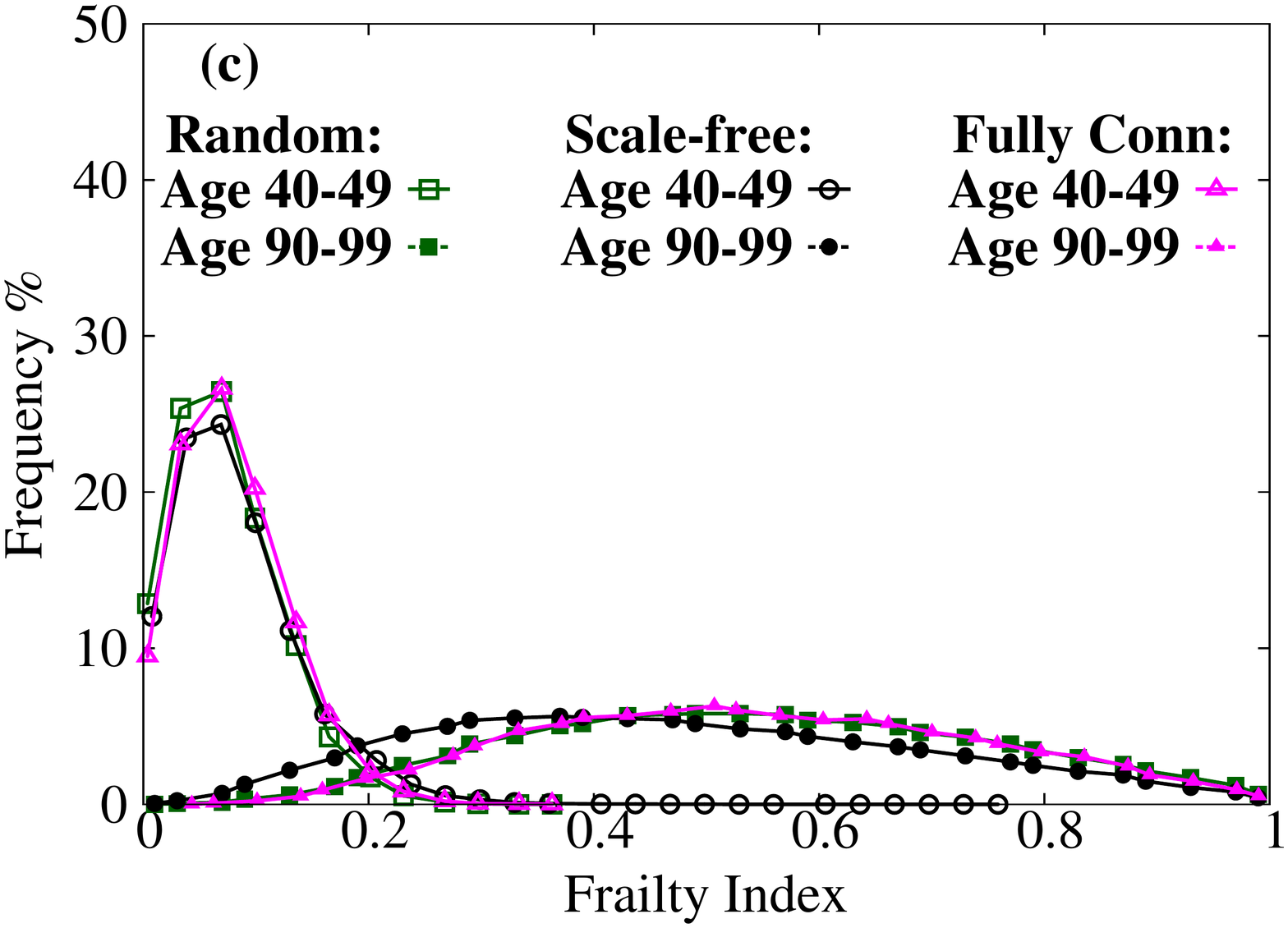}} \\
      \resizebox{70mm}{!}{\includegraphics[trim = 15mm 10mm 20mm 20mm, clip, width=\textwidth]{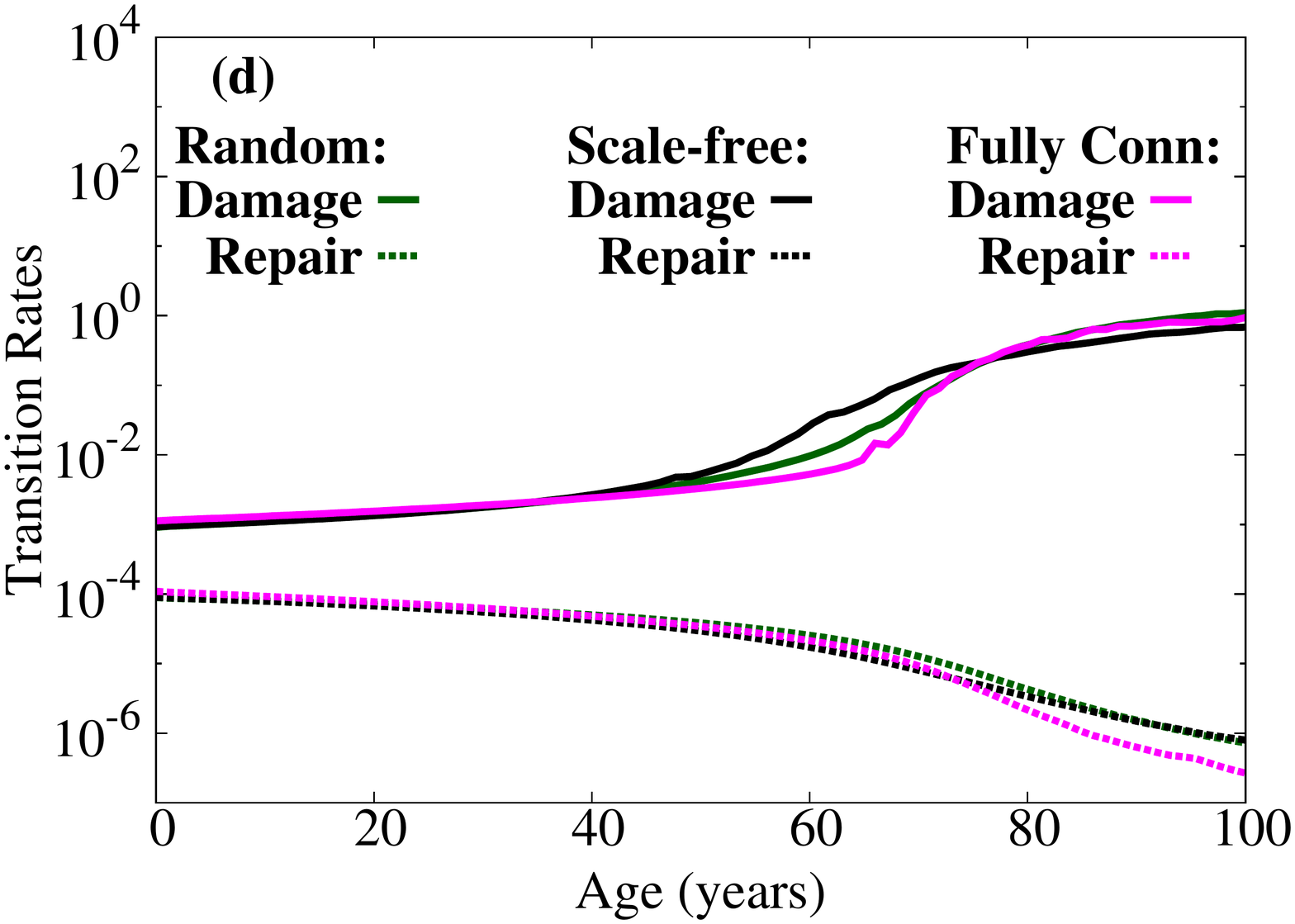}} \\
    \caption{We have done some exploration of different network models. Here we show how random and scale-free networks (both with average degree $20$) compare with a fully connected networks (with degree equal to $N-1$, since nodes don't connect to themselves). (a) mortality rate vs age, (b) $F$ vs age, (c) frailty index distribution $P(F)$ for young (ages $40-49$ years) and old (ages $90-99$ years) cohorts, and (d)  transition rates $\Gamma_\pm$ vs age, averaged over the $F_{30}$ nodes.  We see that while all networks are similar, there are apparent differences in the behavior of $F$ vs age for the scale-free network.  }
    \label{NETWORK}
  \end{center}
\end{figure} 

\begin{figure}[!htbp]
  \begin{center}
      \resizebox{70mm}{!}{\includegraphics[trim = 15mm 10mm 5mm 20mm, clip, width=\textwidth]{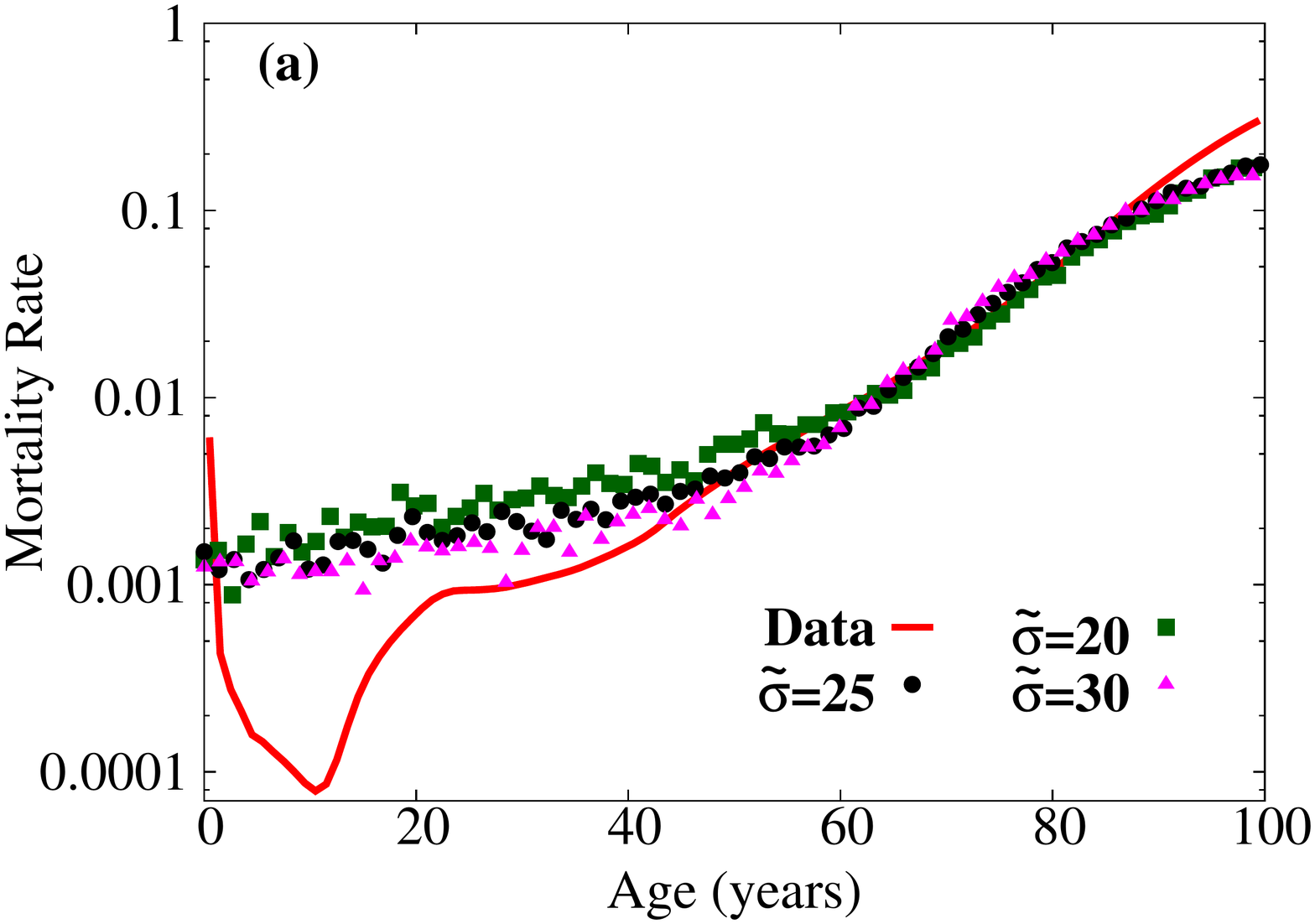}} \\
      \resizebox{70mm}{!}{\includegraphics[trim = 15mm 10mm 5mm 20mm, clip, width=\textwidth]{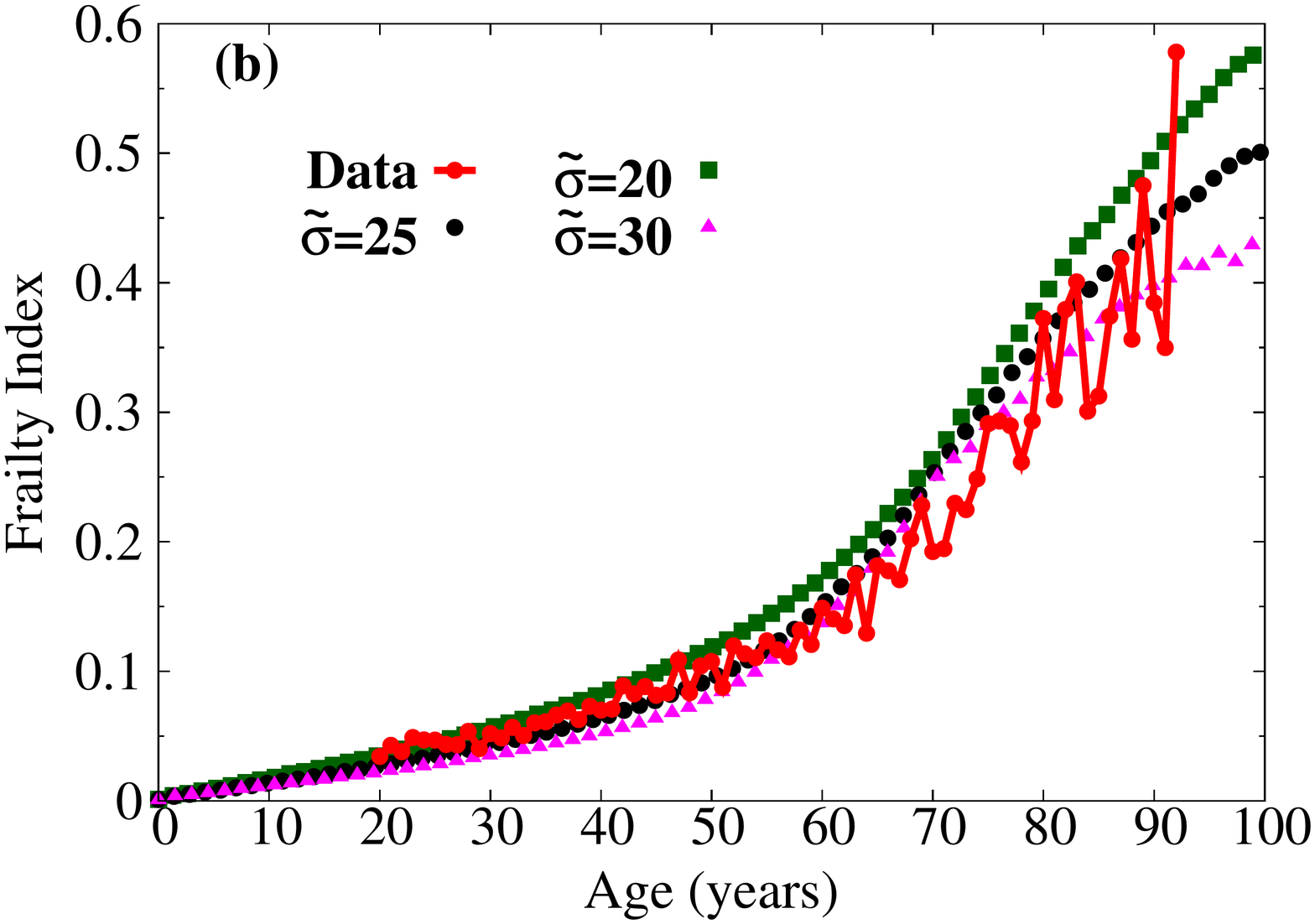}} \\
      \resizebox{70mm}{!}{\includegraphics[trim = 15mm 10mm 20mm 20mm, clip, width=\textwidth]{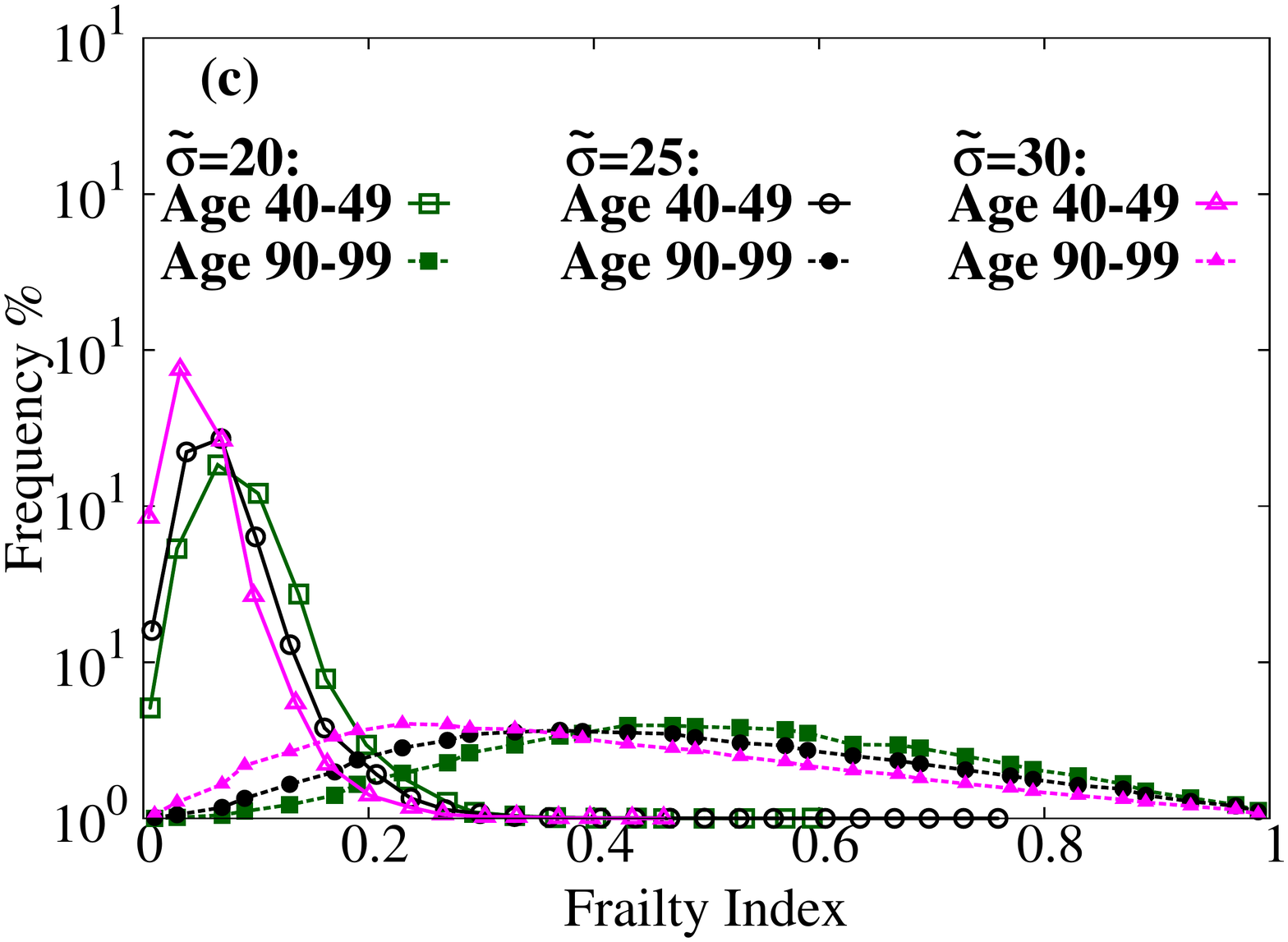}} \\
      \resizebox{70mm}{!}{\includegraphics[trim = 15mm 10mm 20mm 20mm, clip, width=\textwidth]{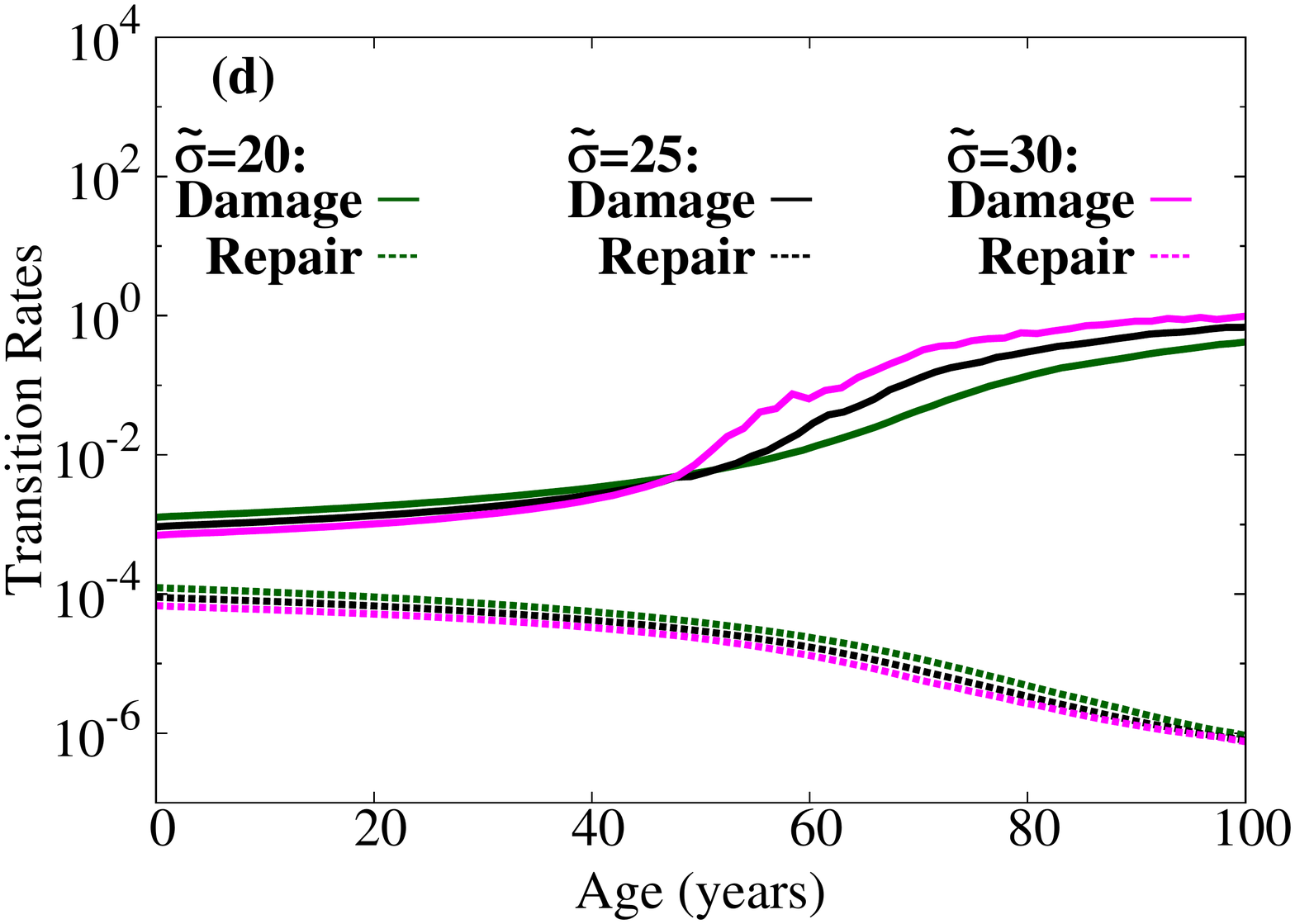}} \\
    \caption{We have varied the scaled interaction $\tilde{\sigma}$, with $\tilde{\sigma}=20$ (green squares), $\tilde{\sigma}=25$ (the default value, black circles), and $\tilde{\sigma}=30$ (pink triangles). (a) mortality rate vs age, (b) $F$ vs age, (c) frailty index distribution $P(F)$ for young (ages $40-49$ years) and old (ages $90-99$ years) cohorts, and (d)  transition rates $\Gamma_\pm$ vs age, averaged over the $F_{30}$ nodes.  We see that the upward curvature of the $F$ vs age plot increases with increasing $\tilde{\sigma}$. }
    \label{SIGMA}
  \end{center}
\end{figure}

\begin{figure}[!htbp]
  \begin{center}
      \resizebox{70mm}{!}{\includegraphics[trim = 15mm 10mm 5mm 20mm, clip, width=\textwidth]{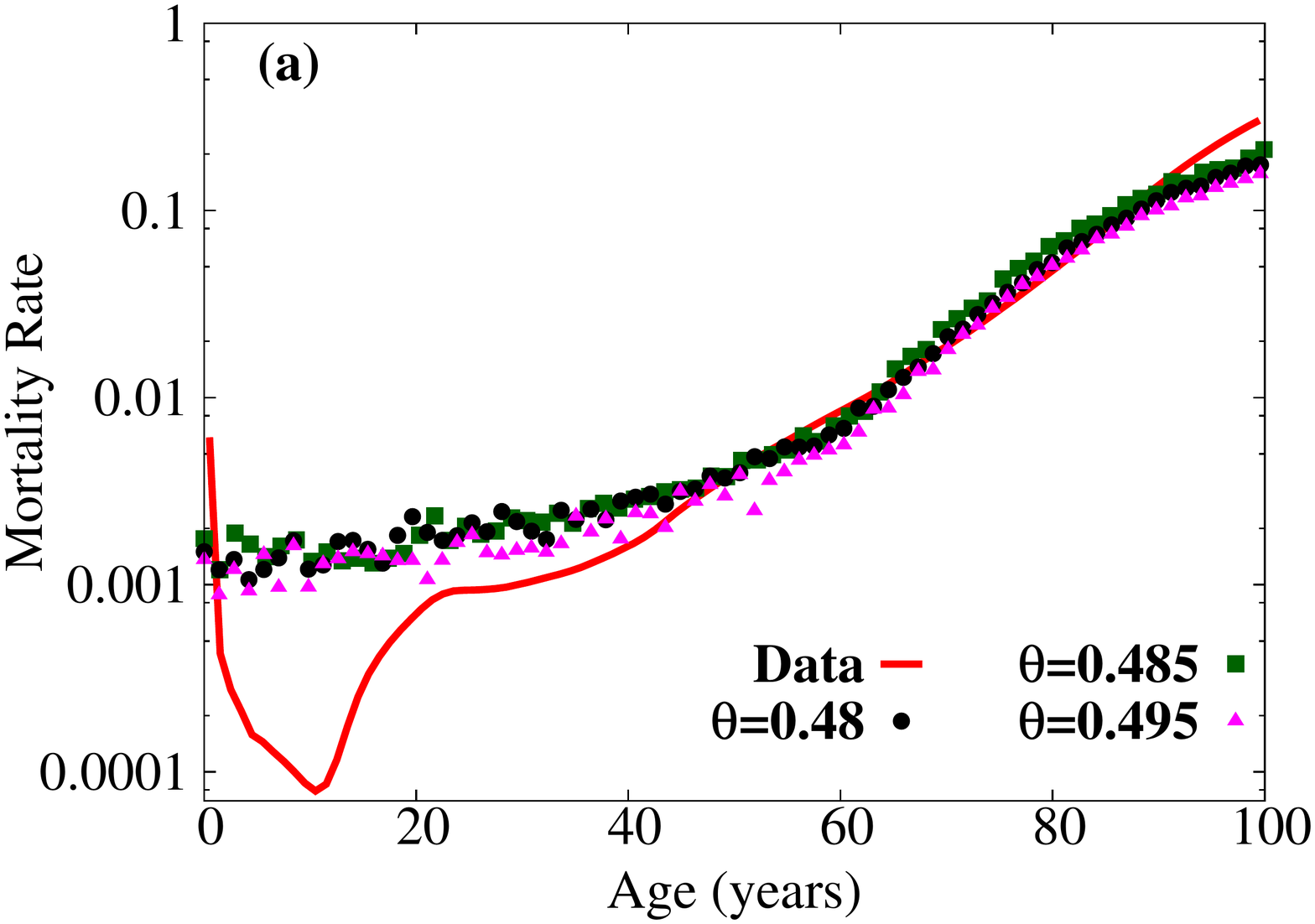}} \\
      \resizebox{70mm}{!}{\includegraphics[trim = 15mm 10mm 5mm 20mm, clip, width=\textwidth]{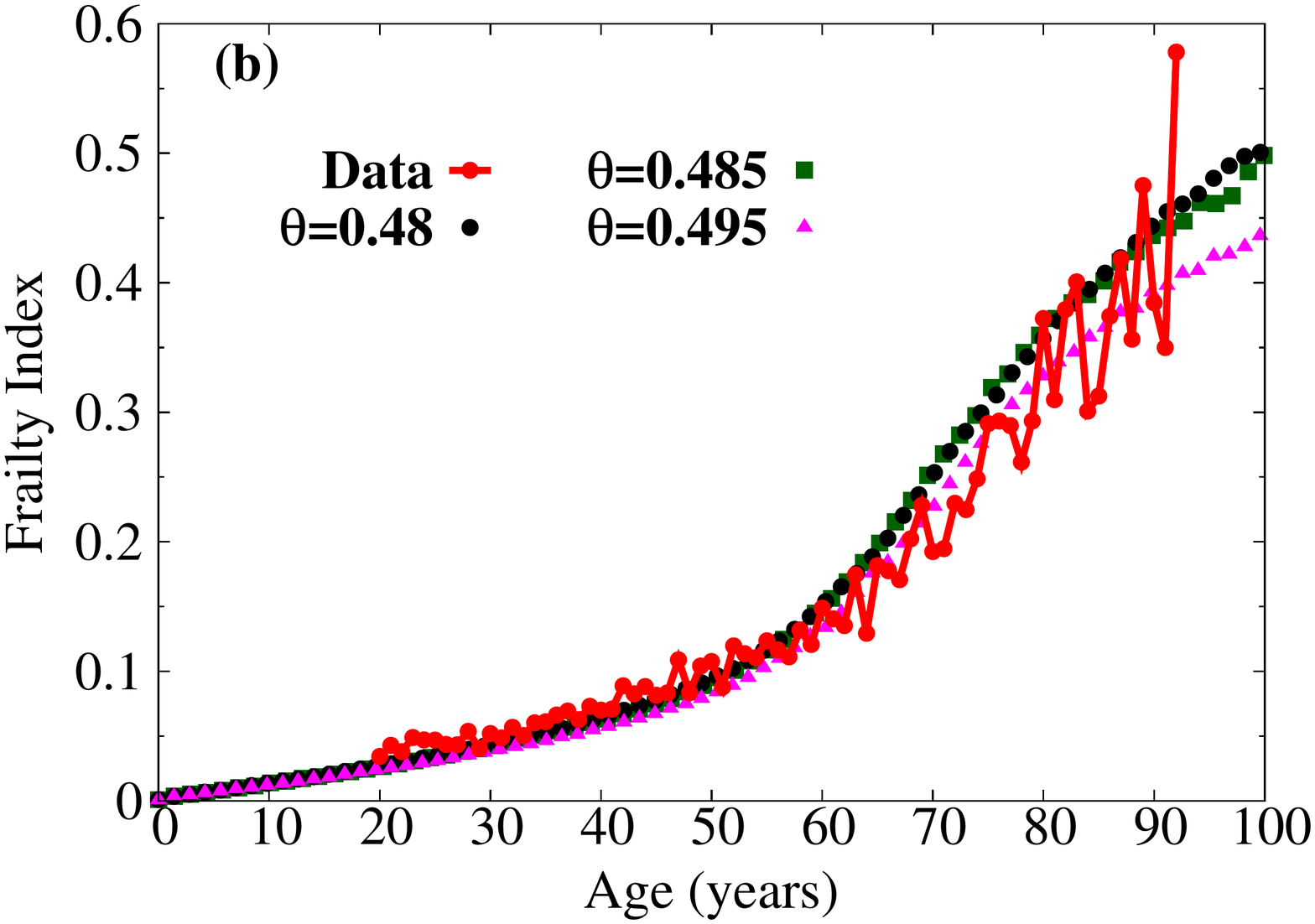}} \\
      \resizebox{70mm}{!}{\includegraphics[trim = 15mm 10mm 20mm 20mm, clip, width=\textwidth]{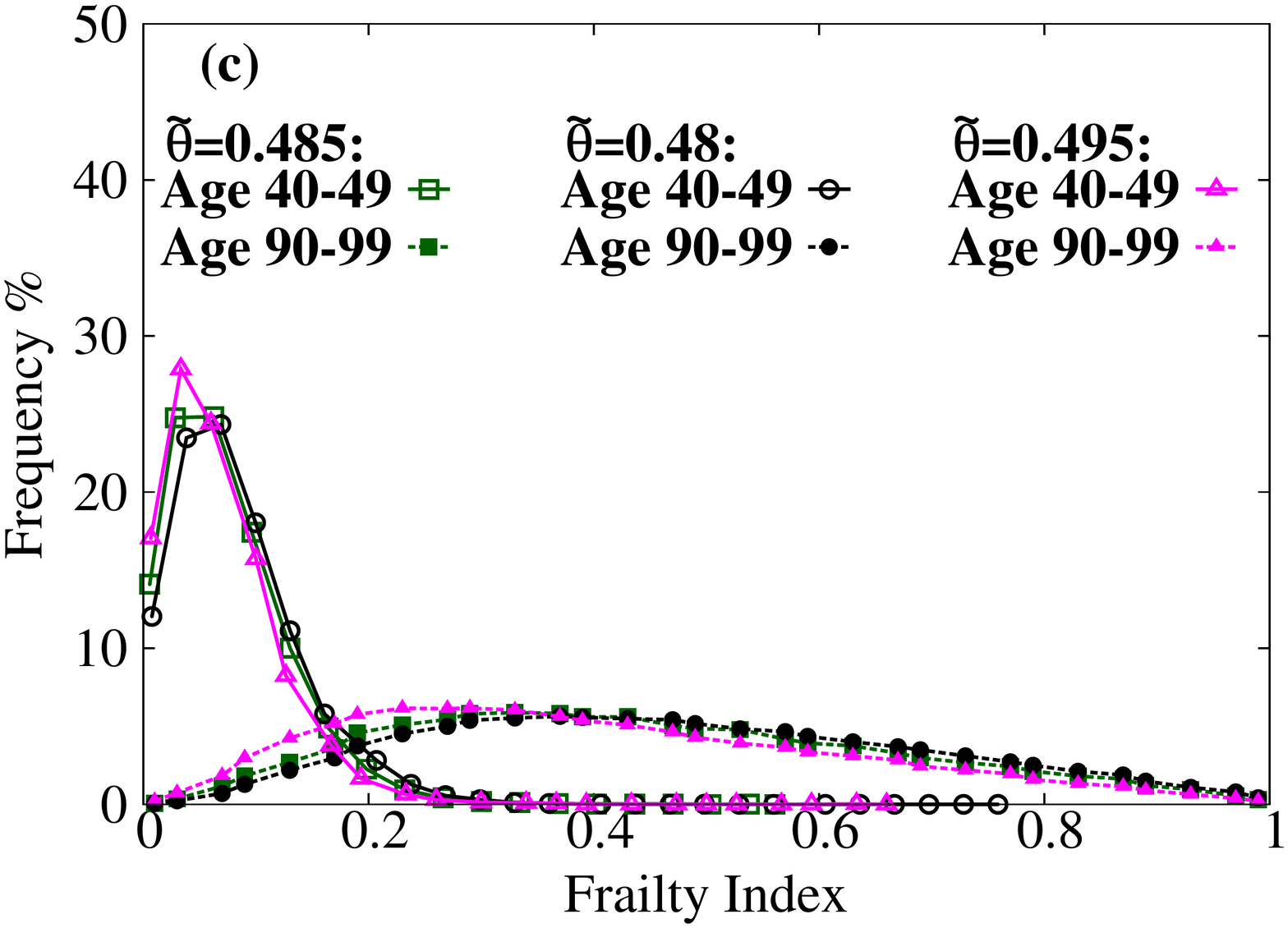}} \\
      \resizebox{70mm}{!}{\includegraphics[trim = 15mm 10mm 20mm 20mm, clip, width=\textwidth]{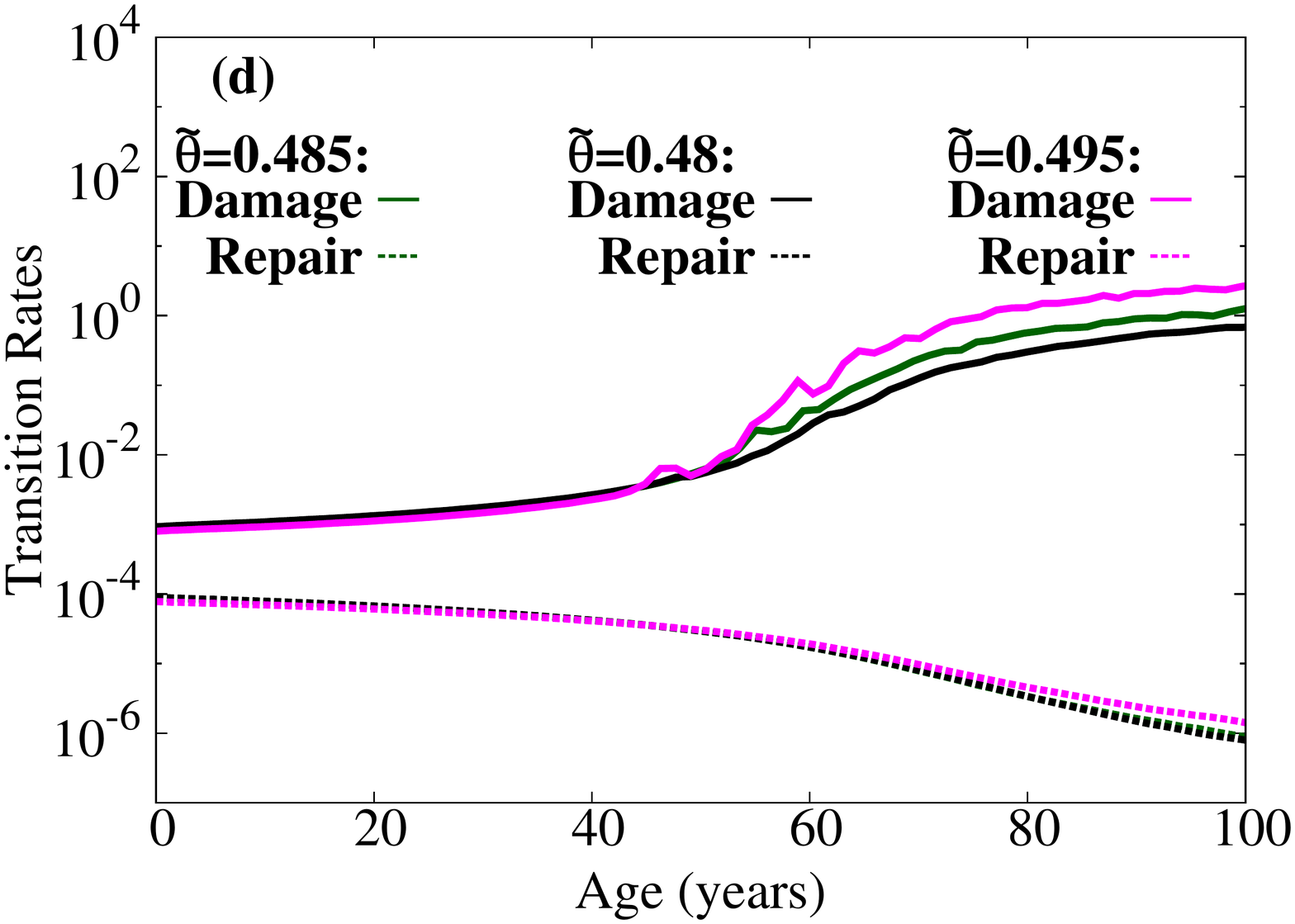}} \\
    \caption{We have varied the shape of the potential, $\theta$, with $\theta=0.48$ (default value, black circles), $\theta=0.485$ (green squares), and $\theta=0.495$ (pink triangles). We note that $\theta=0.5$ corresponds to a symmetric potential, so does not allow asymmetric transition rates at $f_i=0$. (a) mortality rate vs age, (b) $F$ vs age, (c) frailty index distribution $P(F)$ for young (ages $40-49$ years) and old (ages $90-99$ years) cohorts, and (d)  transition rates $\Gamma_\pm$ vs age, averaged over the $F_{30}$ nodes. We see that $\theta$ does not strongly change our results.  }
    \label{THETA}
  \end{center}
\end{figure}

\begin{figure}[!htbp]
  \begin{center}
      \resizebox{70mm}{!}{\includegraphics[trim = 15mm 10mm 5mm 20mm, clip, width=\textwidth]{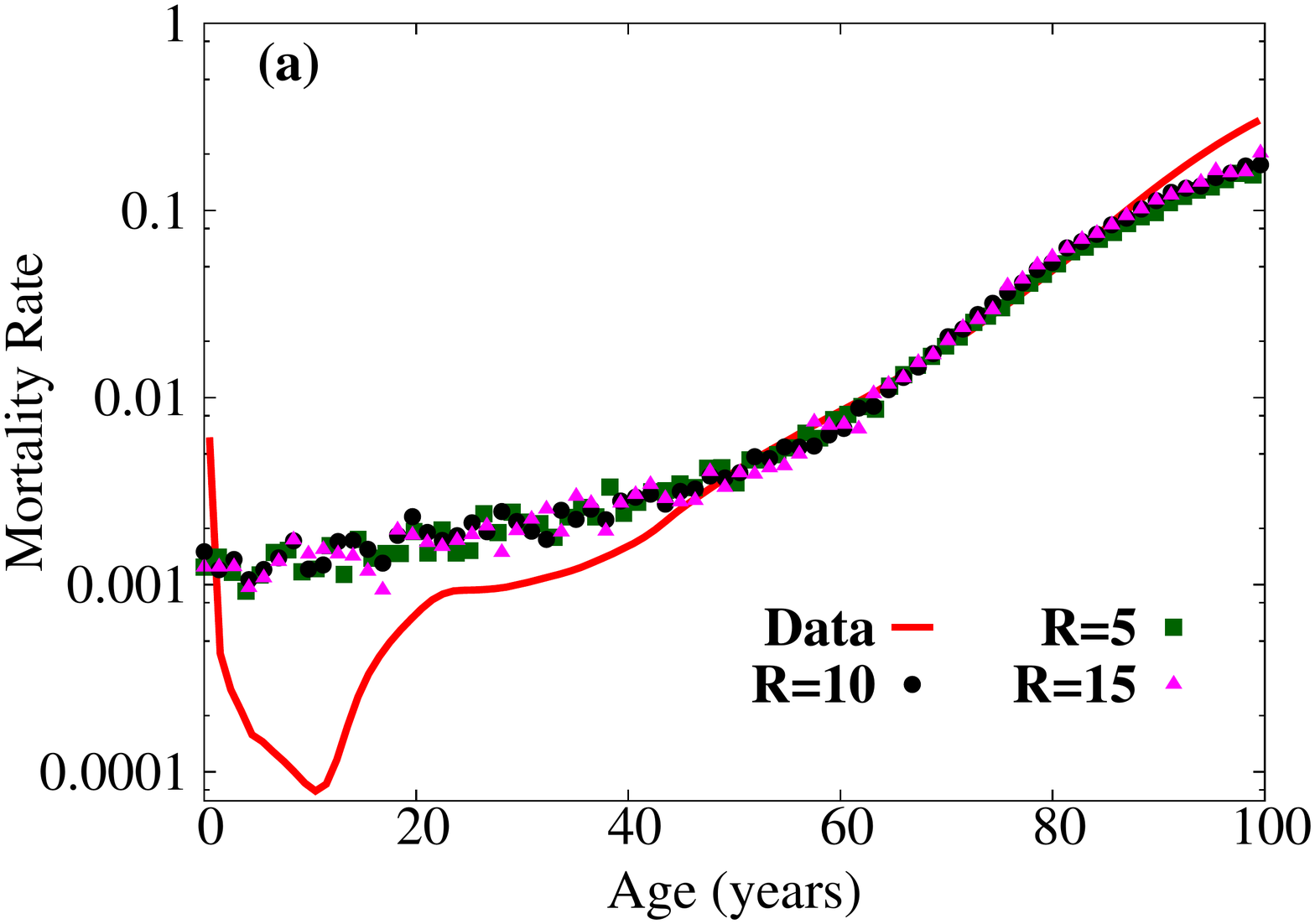}} \\
      \resizebox{70mm}{!}{\includegraphics[trim = 15mm 10mm 5mm 20mm, clip, width=\textwidth]{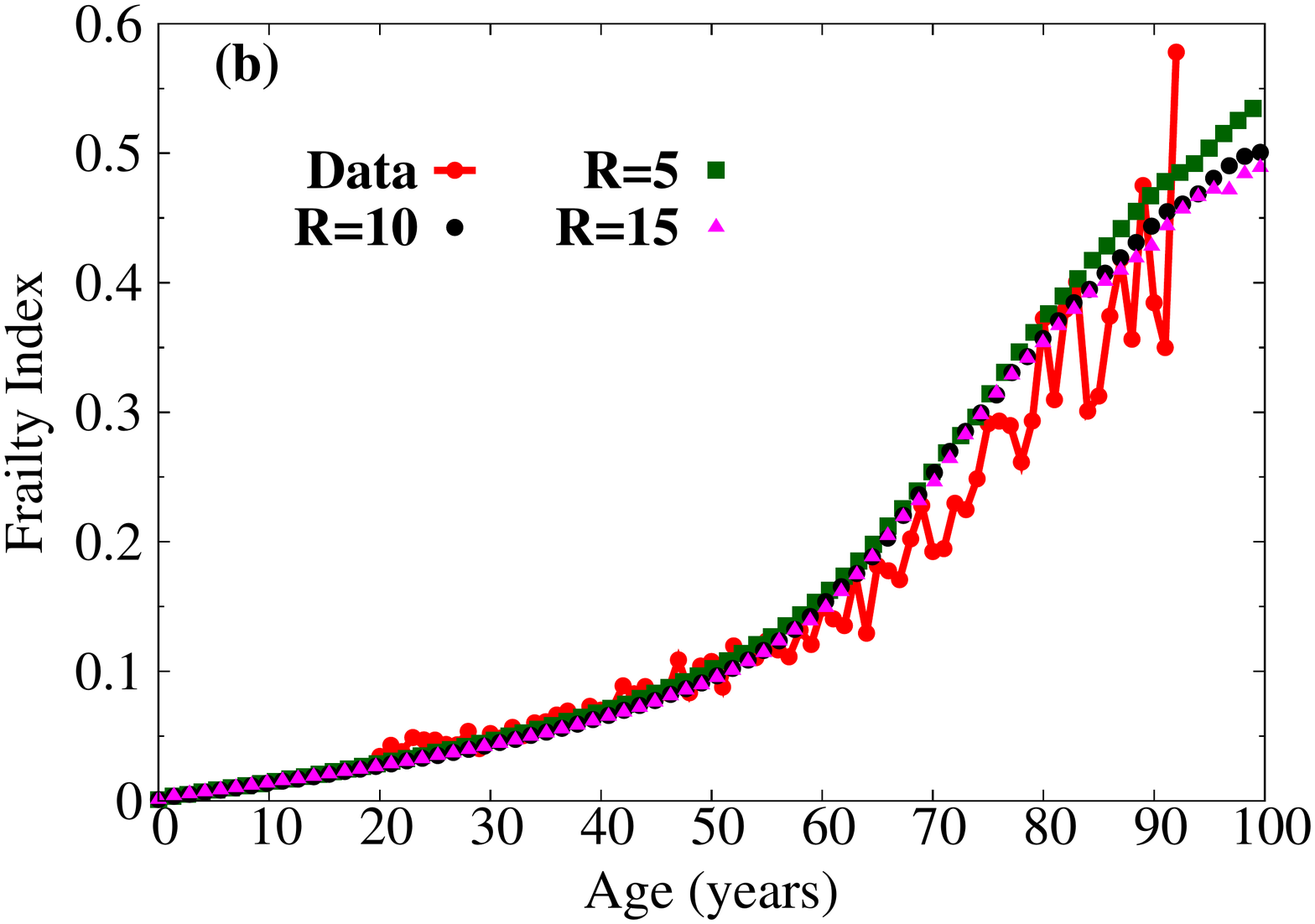}} \\
      \resizebox{70mm}{!}{\includegraphics[trim = 15mm 10mm 20mm 20mm, clip, width=\textwidth]{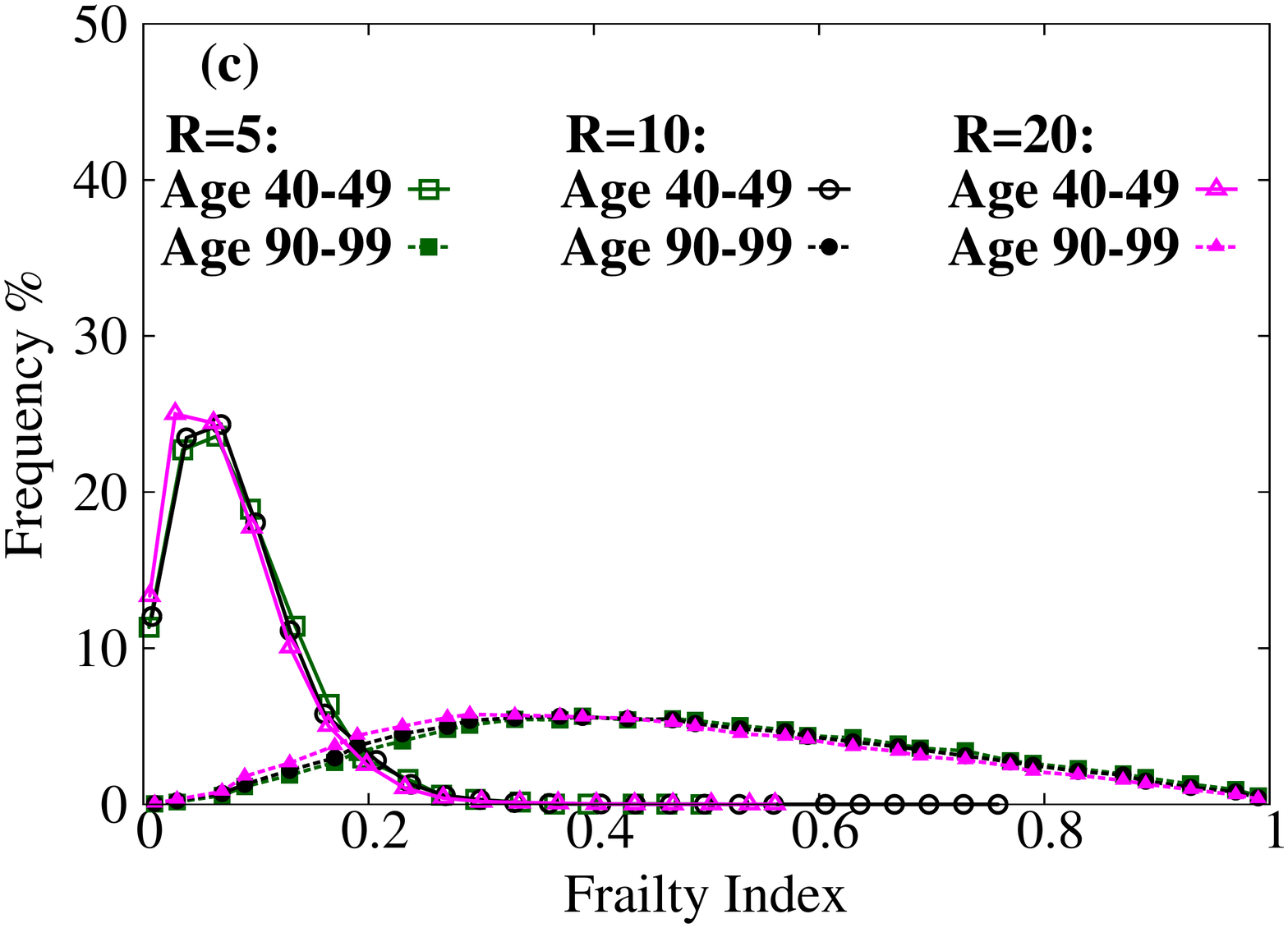}} \\
      \resizebox{70mm}{!}{\includegraphics[trim = 15mm 10mm 20mm 20mm, clip, width=\textwidth]{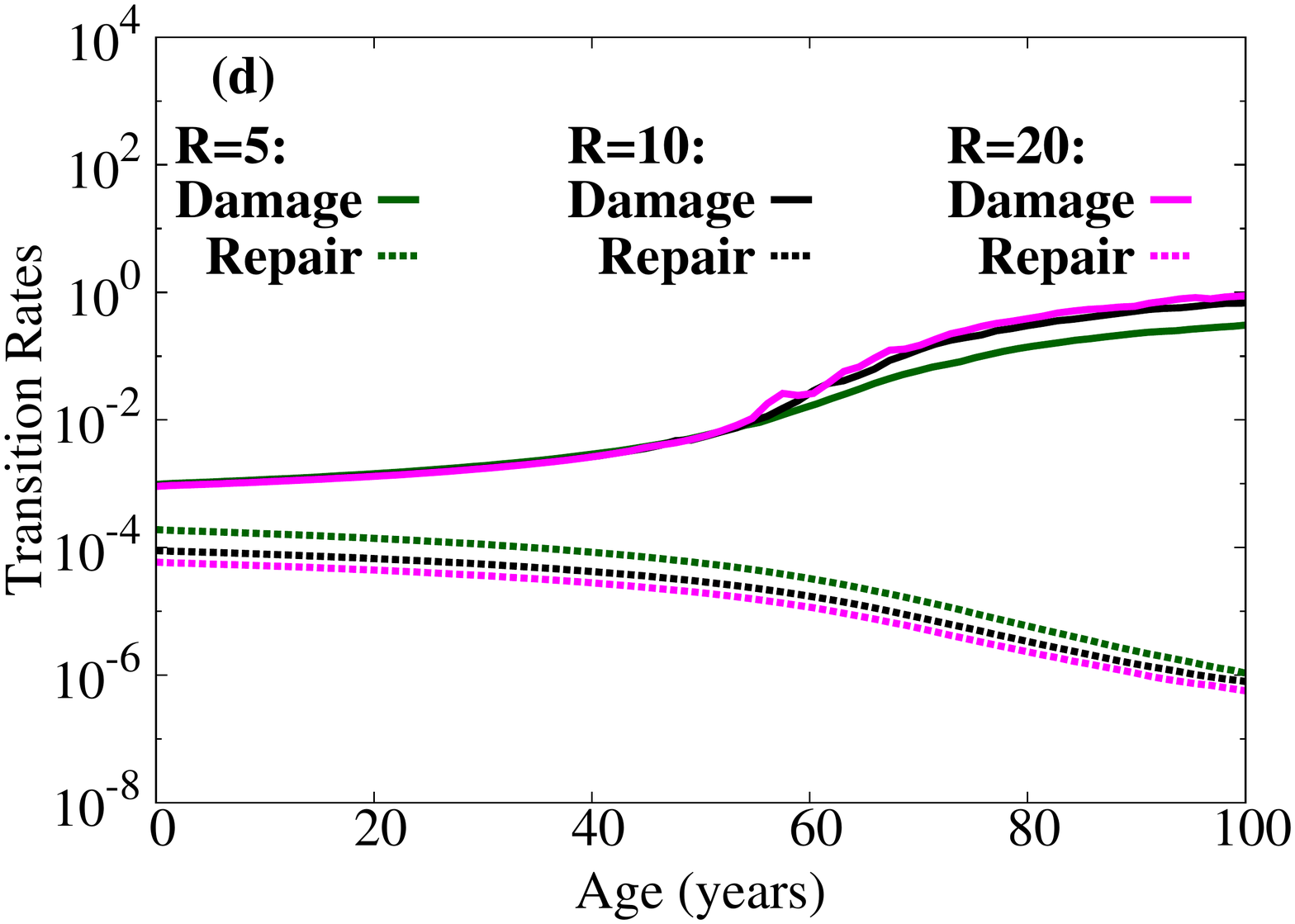}} \\
    \caption{We have varied the initial ratio of damage to repair rates $R \equiv \Gamma_+(0)/\Gamma_-(0)$, with $R=5$ (green triangles), $R=10$ (our default value, black circles), and $R=15$ (pink triangles). (a) mortality rate vs age, (b) $F$ vs age, (c) frailty index distribution $P(F)$ for young (ages $40-49$ years) and old (ages $90-99$ years) cohorts, and (d)  transition rates $\Gamma_\pm$ vs age, averaged over the $F_{30}$ nodes.  We see that $R$ does not strongly change our results.  }
    \label{R}
  \end{center}
\end{figure}

\pagebreak


\end{document}